\documentclass[lettersize,journal]{IEEEtran}
\IEEEoverridecommandlockouts

\usepackage{cite}
\usepackage{amsmath,amssymb,amsfonts}
\usepackage{amssymb}
\usepackage{algpseudocode}
\usepackage{amsfonts}
\usepackage{graphicx}
\usepackage{fancyhdr}  %
\usepackage{cases}
\usepackage{textcomp}
\usepackage{extarrows}
\usepackage{orcidlink}

\usepackage{multirow}
\usepackage{caption}
\usepackage{float} 
\usepackage{subcaption}
\usepackage{makecell}

\usepackage{booktabs}%
\usepackage{array}%
\usepackage{footnote}
\makesavenoteenv{tabular}

\usepackage{algorithm,algpseudocode}
\algnewcommand{\Inputs}[1]{%
  \State \textbf{Inputs:}
  \Statex \hspace*{\algorithmicindent}\parbox[t]{.8\linewidth}{\raggedright #1}
}
\algnewcommand{\Initialize}[1]{%
  \State \textbf{Initialization:}
  \Statex \hspace*{\algorithmicindent}\parbox[t]{.8\linewidth}{\raggedright #1}
}

\usepackage{multirow,tabularx}
\usepackage{multicol,mwe,float,subcaption}
\usepackage{mathtools}
\usepackage{xcolor}
\usepackage[english]{babel}
\usepackage{amsthm}
\usepackage[numbers,sort&compress]{natbib}

\makeatletter
\renewcommand{\maketag@@@}[1]{\hbox{\m@th\normalsize\normalfont#1}}%
\makeatother

\def\BibTeX{{\rm B\kern-.05em{\sc i\kern-.025em b}\kern-.08em
    T\kern-.1667em\lower.7ex\hbox{E}\kern-.125emX}}
\begin{document}

\title{Multi-Sources Fusion Learning for Multi-Points NLOS Localization in OFDM System}

\author{Bohao Wang, Zitao Shuai, Chongwen Huang, Qianqian Yang,
Zhaohui Yang, Richeng Jin, \\ Ahmed Al Hammadi,
Zhaoyang Zhang, Chau~Yuen,~\IEEEmembership{Fellow,~IEEE,} and M\'{e}rouane~Debbah,~\IEEEmembership{Fellow,~IEEE}

\thanks{
  The work was supported in part by the China National Key R\&D Program under Grant 2021YFA1000500 and 2023YFB2904804; in part by the National Natural Science Foundation of China under Grant 62331023, 62101492, 62394292 and U20A20158; in part by the Zhejiang Provincial Key R\&D Program under Grant No. 2023C01021; in part by the Zhejiang Provincial Natural Science Foundation of China under Grant LR22F010002; in part by the Zhejiang Provincial Science and Technology Plan Project under Grant 2024C01033, in part by the Zhejiang University Global Partnership Fund; in part by the Fundamental Research Funds for the Central Universities under Grant No. 226-2024-00069; and in part by the Ministry of Education (MOE), Singapore, through its MOE Tier 2 (Award number MOE-T2EP50220-0019) and A\*STAR (Agency for Science, Technology and Research) Singapore, under Grant No. M22L1b0110. \textit{(Corresponding author: Chongwen Huang.)}}

\thanks{B. Wang, C. Huang are with the College of Information Science and Electronic Engineering, Zhejiang University, Hangzhou 310007, China, with the State
Key Laboratory of Integrated Service Networks, Xidian University, Xi’an 710071, China, and with Zhejiang Provincial Key Laboratory of Info. Proc., Commun. \& Netw. (IPCAN), Hangzhou 310027, China. (E-mails: \{bohaowang, chongwenhuang\}@zju.edu.cn).}

\thanks{Z. Shuai is with the Department of Electrical Engineering and Computer Science, University of Michigan, Ann Arbor, MI 48109-2122. (E-mail: ztshuai@umich.edu).}

\thanks{Q. Yang, Z. Yang, R. Jin, Z. Zhang are with the College of Information Science and Electronic Engineering, Zhejiang University, Hangzhou 310007, China, and also with Zhejiang Provincial Key Laboratory of Info. Proc., Commun. \& Netw. (IPCAN), Hangzhou 310027, China. (E-mails: \{qianqianyang20, yang\_zhaohui, richengjin, ning\_ming\}@zju.edu.cn).}

\thanks{A. Al Hammadi is with the Technology Innovation Institute, 9639 Masdar City, AbuDhabi, United Arab Emirates. (E-mail: Ahmed.Alhammadi@tii.ae).}

\thanks{C. Yuen is with the School of Electrical and Electronics Engineering, Nanyang Technological University, Singapore 639798, Singapore. (E-mail: chau.yuen@ntu.edu.sg).}

\thanks{M. Debbah is with KU 6G Research Center, Khalifa University of Science and Technology, PO Box 127788, Abu Dhabi, United Arab Emirates. (E-mail: merouane.debbah@ku.ac.ae).}
\vspace{-8mm}

}

\maketitle

\begin{abstract}

Accurate localization of mobile terminals is a pivotal aspect of integrated sensing and communication systems. 
Traditional fingerprint-based localization methods, which infer coordinates from channel information within pre-set rectangular areas, often face challenges due to the heterogeneous distribution of fingerprints inherent in non-line-of-sight (NLOS) scenarios, particularly within orthogonal frequency division multiplexing systems.
To overcome this limitation, we develop a novel multi-sources information fusion learning framework referred to as the Autosync Multi-Domains NLOS Localization (AMDNLoc). 
Specifically, AMDNLoc employs a two-stage matched filter fused with a target tracking algorithm and iterative centroid-based clustering to automatically and irregularly segment NLOS regions, ensuring uniform distribution within channel state information across frequency, power, and time-delay domains. 
Additionally, the framework utilizes a segment-specific linear classifier array, coupled with deep residual network-based feature extraction and fusion, to establish the correlation function between fingerprint features and coordinates within these regions. 
Simulation results reveal that AMDNLoc achieves an impressive NLOS localization accuracy of 1.46 meters on typical wireless artificial intelligence research datasets and demonstrates significant improvements in interpretability, adaptability, and scalability.

\end{abstract}

\begin{IEEEkeywords}
Multi-sources, information fusion, fingerprint localization, inverse, heterogeneity, regional covariant.
\end{IEEEkeywords}

\section{Introduction}\label{sec:intro}
As a key usage scenario for sixth-generation communications, integrated sensing and communication requires a low-latency and high-precision mobile terminal (MT) localization, especially in fields like the smart city \cite{palmieri2016cloud}, internet of vehicles \cite{kaiwartya2016internet}, telemedicine \cite{perednia1995telemedicine} and so on. 
In dense urban environments, the localization accuracy of traditional global navigation satellite systems (GNSS) and ranging-based localization techniques is significantly limited in the multi-points non-line-of-sight (NLOS) scenarios \cite{hsu2018analysis, karttunen2017spatially}. 
Fortunately, systems related to orthogonal frequency division multiplexing (OFDM) technology, such as sparse code multiple access-OFDM and OFDM access, present a novel opportunity for accurate NLOS localization. These systems enable the unique characterization of each point within a scene through distinct multi-path features utilizing orthogonal frequency division techniques, thereby enhancing the efficacy of fingerprint-based localization methods.

Taking channel state information (CSI) associated with multi-path features as fingerprints, the fingerprint-based localization method is to estimate locations by matching real-time measurements with pre-stored fingerprints, offering advantages in terms of lower energy consumption and implementation cost \cite{csik2023fingerprinting, 10137350, 8920098}.
However, this process is particularly challenging in outdoor multi-points NLOS scenarios. The complexity arises from the influence of various urban elements, such as buildings and scattered objects, which induce substantial heterogeneity in the fingerprint distribution. 
This heterogeneity leads to an uneven regional correlation between fingerprints and their corresponding locations, thereby making it difficult to deduce locations from fingerprints \cite{xiao2018learning}.
 
\begin{figure*}\vspace{-0mm}
	\begin{center}
		\centerline{\includegraphics[width=0.95\textwidth]{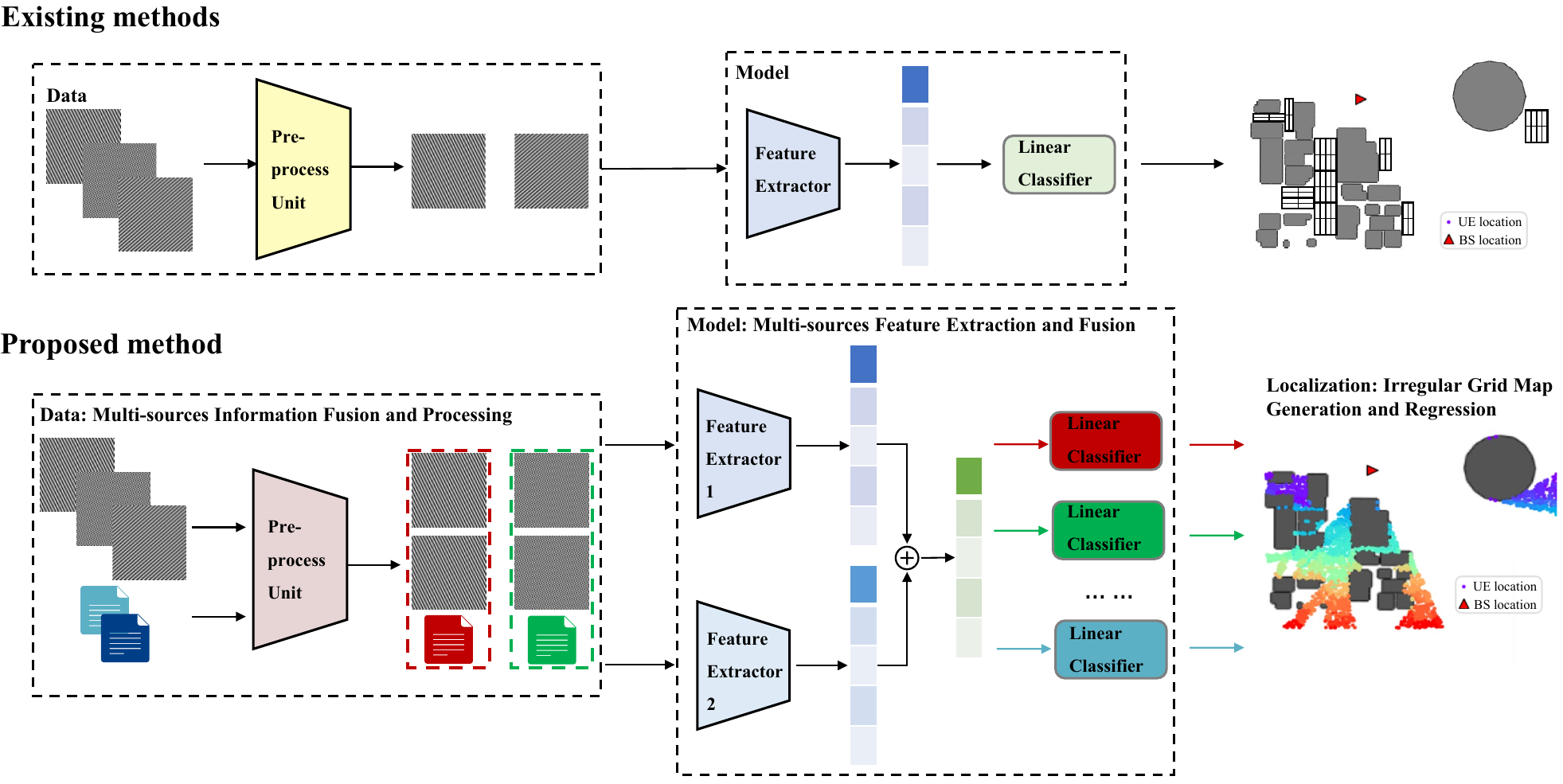}}  \vspace{-0mm}
		\caption{System diagram of proposed AMDNLoc framework compared with existing methods.}
		\label{fig:framework} \vspace{-10mm}
	\end{center}
\end{figure*}

To tackle the issue, conventional approaches often involve partitioning the coverage area into multiple cells, typically arranged in a grid pattern. Each cell in this grid map is associated with a reference fingerprint for localization purposes.
Presently, a considerable number of research focuses on regular grid maps \cite{liu2007survey, vo2015survey, sun2018single, peng2019decentralized, sun2019fingerprint, li2020deep, del2019localization, gante2020deep, gong2023transformer}. This approach entails segmenting a large area into medium-sized square cells based on geographical coordinates. All the fingerprints in each cell is labeled uniformly, and the cell size is iteratively reduced until the closest fingerprint in the dataset is matched. 
While it's easy to manually divide regular areas, localization accuracy is highly sensitive to the specifics of the regional divisions, such as cell size and alignment.
To mitigate the problem, researchers have proposed irregular grid maps, which feature variably distributed cells. 
For example, \cite{bittencourt2018proposal} adapts the grid to align with the urban road network,  anticipating that outdoor mobile users will predominantly follow these paths. 
\cite{kim2014design} introduces clustered merging, grouping geographically close fingerprints that are statistically significant for large-scale Wi-Fi positioning systems. 
Additionally, \cite{bittencourt2023generation} employs the farthest-first traversal algorithm and low-discrepancy sequences to generate irregular grid maps of received signal strength (RSS).
Irregular grid maps can potentially enhance performance and location estimation precision. However, the methods for generating these maps are computationally intensive. Moreover, due to the heterogeneity issue, despite the tendency for fingerprints at adjacent positions to exhibit high similarity, existing works do not guarantee that all fingerprints within a cell, whether regular or irregular, will share this level of similarity.


Building upon the foundation of manually divided regular grid maps, various artificial intelligent (AI) techniques have been developed to establish an inverse correlation function between fingerprints and locations \cite{peng2019decentralized, sun2019fingerprint, li2020deep, del2019localization, gante2020deep, gong2023transformer, ferrand2020dnn}.
For instance, \cite{li2020deep} utilizes a deep residual network (ResNet) for coarse localization, complemented by multi-layers perceptron-based transfer learning for fine tuning. 
\cite{sun2019fingerprint} introduces a CALP deep convolutional neural network (DCNN), which hierarchically refines the search area to pinpoint the closest fingerprint. 
Furthermore, \cite{gong2023transformer} presents a deep neural network architecture incorporating the Transformer model, relying on self-attention mechanisms to process sequences of fingerprint patches.
While AI techniques have significantly enhanced localization accuracy, they also introduce complexities. The need for continuous mapping of real-time measurements to smaller square grid cells entails updating numerous parameters. Additionally, if the initial coarse positioning of a fingerprint is inaccurately predicted, it can lead to cumulative errors in the final results. 
Moreover, despite the unique multi-path features of each location, existing approaches may not fully capture this distinctiveness, since they often rely on a single type of fingerprint—such as RSS, channel impulse response (CIR), channel frequency response (CFR), the angle-delay channel amplitude matrix (ADCAM)—among others.
Experimental evidence indicates that in various datasets, a proportion of remote points share strikingly similar fingerprints within a scene, challenging the foundational assumption of independent and identically distributed (i.i.d) data. This similarity can significantly undermine the accuracy of the predicted results \cite{kanhere2021position, yang2013rssi, bhattacherjee2020localization}. Such findings suggest the need for further refinement in these AI-driven localization methods to account for the complexity and diversity of real-world environments.

To address above challenges in large-scale outdoor multi-points NLOS localization, we propose a novel multi-sources information fusion learning framework named the autosync multi-domains NLOS
localization (AMDNLoc). This framework, as depicted in Fig. \ref{fig:framework}, is innovative in its automatic and irregular generation of classification regions, utilizing a fusion of frequency, power, and time-delay information from CSI for the first time. The key contributions are as follows:
\begin{itemize}
    \item We have developed a two-stage matched filter that identifies and extracts all parallel feature intervals related to CFR distribution, named parallel feature CFR (PFCFR). This filter classifies all CFR images according to the closest PFCFR using a target tracking algorithm.
    \item We employ iterative centroid-based clustering to partition the ADCAM, considering power, angle, and delay characteristics. This process combines the classification areas from both the PFCFR and ADCAM, ensuring uniform fingerprint distribution characteristics.
    \item We remove the special samples through data cleaning, and introduce a segment-specific linear classifier array, which, coupled with a deep residual network for feature extraction and fusion, inversely maps fingerprints to coordinates in respective regions while maintaining global features of the complete scene.
    \item The AMDNLoc framework has achieved state-of-the-art (SOTA) results in both ray-tracing wireless AI research dataset (WAIR-D) \cite{huangfu2022wair} and DeepMIMO dataset \cite{alkhateeb2019deepmimo}. Additionally, we have visualized classification areas and compared multiple processing approaches, confirming the enhanced adaptability and scalability of AMDNLoc. The full code of this paper is provided on \cite{gitAMDNloc} as well.
\end{itemize}

The rest of the paper is organized as follows: Section II discusses the channel model and formulates the heterogeneity problem. Section III delves into the AMDNLoc framework, followed by numerical results in Section IV. We conclude our findings in Section V.

Lastly, specific notations are used throughout this paper for clarity. Scalars, vectors, and matrices are represented by $a$, $\mathbf{a}$, and $\mathbf{A}$, respectively. Operations on matrices such as transpose ($\mathbf{A}^T$), Hermitian (conjugate transpose, $\mathbf{A}^H$), inverse ($\mathbf{A}^{-1}$), pseudo-inverse ($\mathbf{A^\dag}$), and Frobenius norm ($\|\mathbf{A}\|_F$) are denoted as shown. The $(m,n)$-th entry of a matrix is indicated by $a{mn}$, and $|\cdot|$ denotes the modulus. The notation $diag(\mathbf{a})$ and $Tr(\mathbf{a})$ signify the diagonal matrix and trace of $\mathbf{a}$, respectively, with $o$ representing infinitesimal of higher order.

\section{System Model}\label{sec:sys}

In this section, we begin by presenting the channel model, and subsequently formulate the heterogeneity problem from the NLOS fingerprint localization perspective.

\begin{figure}[ht]
	\begin{center}
		\centerline{\includegraphics[width=0.475\textwidth]{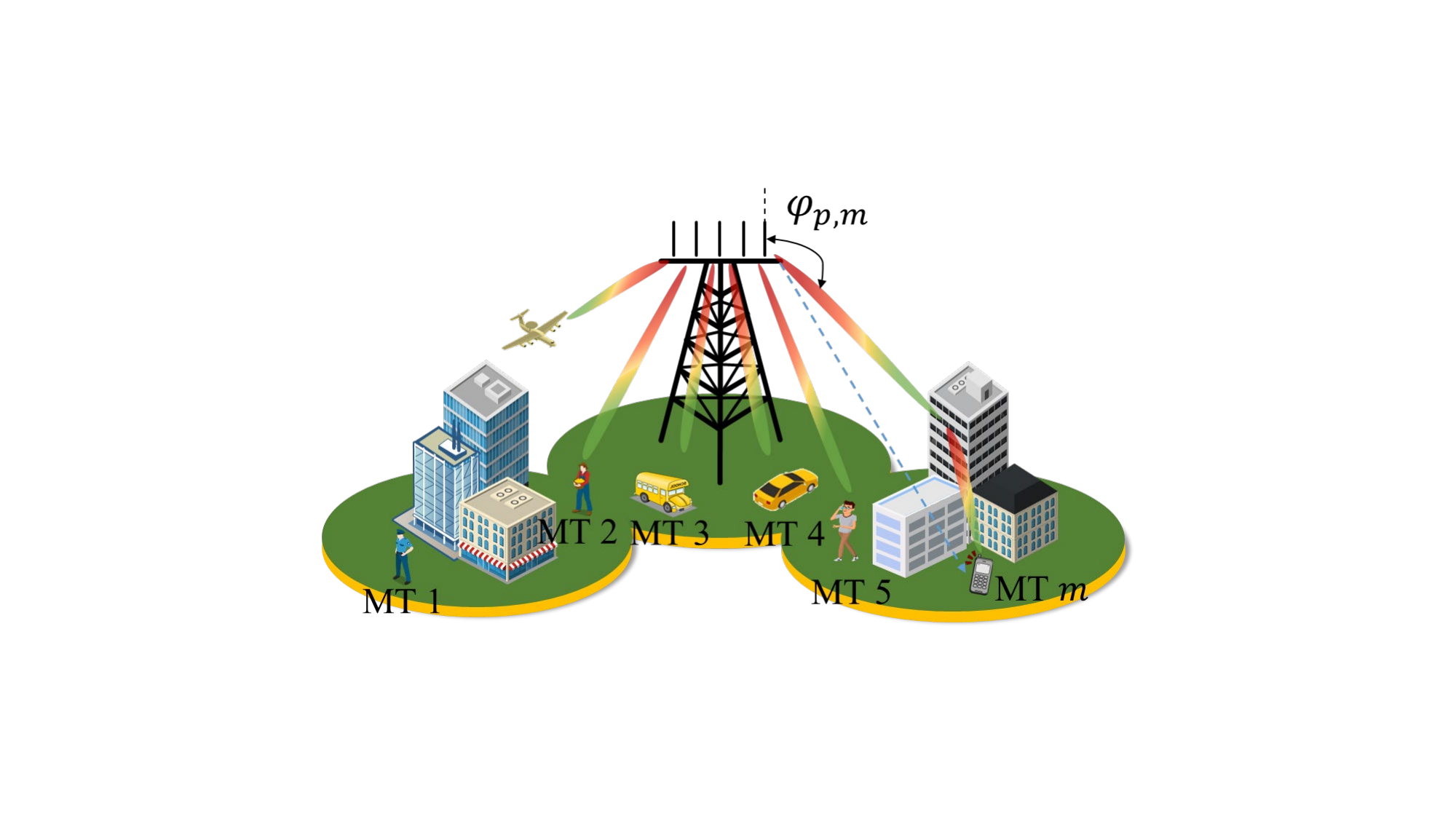}}
		\caption{The considered multiple MTs systems equipped with OFDM technology. }
		\label{fig:system}
		\vspace{-10mm}
	\end{center}
\end{figure}

The considered downlink communication system of multiple MTs equipped with OFDM technology is illustrated in Fig. \ref{fig:system}. The BS is equipped with a uniform linear array (ULA) comprising $N_{t}$ antennas. We have $m\in\{1, ..., M\}$ MTs, each equipped with a single omni-directional antenna. The angle of arrival (AOA) and the physical distance between the transmit antenna and the first receive antenna associated with the $p$-th path are denoted by $\phi_{p,m}\in(0,\pi)$ and $d_{p,m}$, respectively.

The CIR vector of the $m$-th user is given by:
\begin{equation}
{{\bf q}}_{p,m} = a_{p,m}{\bf e}\left(\phi_{p,m}\right),
\end{equation}
where $a_{p,m}\sim\mathcal CN(0,\sigma_{p,m})$ denotes the complex gain of the $p$-th path, and ${\bf e}\left(\phi\right)$ is the array response vector corresponding to the AOA $\phi$. Eq. It represents the unified form for calculating the CIR for all paths, irrespective of whether they are LOS or NLOS. 
It takes the form:
\begin{equation}
{{\bf e}}\left(\phi \right) = \left[1,{e^{ - {\bar{\jmath }}2\pi { {{d\cos \left(\phi \right)} \over {{\lambda _c}}}}}}, \ldots,{e^{ - {\bar{\jmath }}2\pi { {{\left({{N_t} - 1} \right)d\cos \left(\phi \right)} \over {{\lambda _c}}}}}} \right]^T,
\end{equation}
where $j=\sqrt{-1}$, $d$ is the antenna spacing (typically $\lambda/2$ in recent MIMO systems); $\tau_{p,m}=n_{p,m}T_s$ represents the distinguishable propagation delay associated with the $p$-th path; $n_{p,m}$ refers to the sampled delay for the $p$-th path, and $T_s$ denotes the sample interval. The CFR is defined as the sum of time-domain CIRs, each having unique delays:
\begin{equation} {{{\bf h}}_{m,l}} = \sum \limits _{p = 1}^P {{a_{p,m}}{{\bf e}}\left({{\phi _{p,m}}} \right){e^{ - {\bar{\jmath }}2\pi { {{l{n_{p,m}}} \over {{N_c}}}}}}},  \end{equation}
where $N_c$ is the number of subcarriers in the OFDM system, $l\in \{1,...,N_c\}$ . Then, the overall CFR matrix known to the BS can be denoted as the stack of ${{{\bf h}}_{m,l}}$, i.e.,
\begin{equation} \label{H}{{{\bf H}}_m} = \left[ {{{{\bf h}}_{m,0}},{{{\bf h}}_{m,1}}, \ldots,{{{\bf h}}_{m,{N_c} - 1}}} \right]. \end{equation}

Then, authors in \cite{sun2018single} established a mapping from the CFR matrix to a sparse structure by employing DFT operations, referred to as the angle-delay channel response matrix.

Let us define the DFT matrix ${\mathbf{V}}\in \mathbb{C}^{N_t\times N_t}$ as:
\begin{equation*}
{[{\mathbf{V}}]_{z,q}} \triangleq \frac{1}{{\sqrt {{N_t}} }}{e^{ - j2\pi \frac{{\left( {z\left( {q - \frac{{{N_t}}}{2}} \right)} \right)}}{{{N_t}}}}},
\end{equation*}
and ${\mathbf{F}}\in \mathbb{C}^{N_c\times N_c}$ as:
\begin{equation*}
{[{\mathbf{F}}]_{z,q}} \triangleq \frac{1}{{\sqrt {{N_c}} }}{e^{ - j2\pi \frac{{zq}}{{{N_c}}}}},
\end{equation*}
Therefore, the ADCAM of user $m$ is defined as:
\begin{equation}\label{A}
{\left[ {{\bf {A}_m}} \right]_{z,q}} = \mathbb {E}\left\lbrace {\left| {{\left[{{{\mathbf{V}}^H}{\mathbf{H}_m\mathbf{F}}}\right]_{z,q}}} \right|} \right\rbrace,
\end{equation}
where the $(z,q)$ element of the ADCAM represents the absolute gain of the $z$-th AOA and $q$-th delay of the channel.

As we all know, the fingerprint is determined by locations and a specific configuration of the propagation environment:
\begin{equation}
\textit{Fingerprint} = \boldsymbol{\mathcal F}(\boldsymbol{p}, pv),
\end{equation}
where $\textit{Fingerprint}$ represents the channel characteristics used for localization. $\boldsymbol{p} := [x, y]$ denotes the 2D positional coordinates of the users, and $pv$ encompasses the system parameters which remain constant during training and testing, such as frequency, antenna response, and so on.

Traditional fingerprint localization aims to inversely map the $\textit{Fingerprint}$ back to $\boldsymbol{p}$.
However, a great number of devices access the network by non-orthogonal way usually poses challenges to localization accuracy. The impact of different buildings on various areas differs, so that changes in the fingerprint in different areas play different roles in affecting user locations. This leads to the issue of fingerprint heterogeneity in distribution.
Therefore, the inverse process of fingerprint extraction should be expressed as:
\begin{equation}
\hat{\boldsymbol{p}}_{\text{real}} = \boldsymbol{\mathcal F^{-1}}(\textit{Fingerprint}, pv, cv),
\end{equation}
where $cv$ represents a covariate related to the region, weather, and other factors. Our objective in this research is to eliminate the regional covariate and establish a one-to-one relationship solely between $\boldsymbol{p}$ and $\textit{Fingerprint}$.

\section{AMDNLoc Multi-sources Framework}\label{sec:sys}
In this section, we delve into the heterogeneity matched filter, the network structure design, and the detailed network training methodology within the AMDNLoc framework.

\subsection{Design and Application of a Heterogeneity Matched Filter}
In tackling the complex issue of heterogeneous fingerprint distribution, our research has led to the development of a specialized classification approach to uniquely address the distinct distribution characteristics inherent to both CFR and ADCAM.

\subsubsection{PFCFR}
\begin{figure}[htbp]
	\centering
	\begin{subfigure}{0.325\linewidth}
		\centering
		\includegraphics[width=0.9\linewidth]{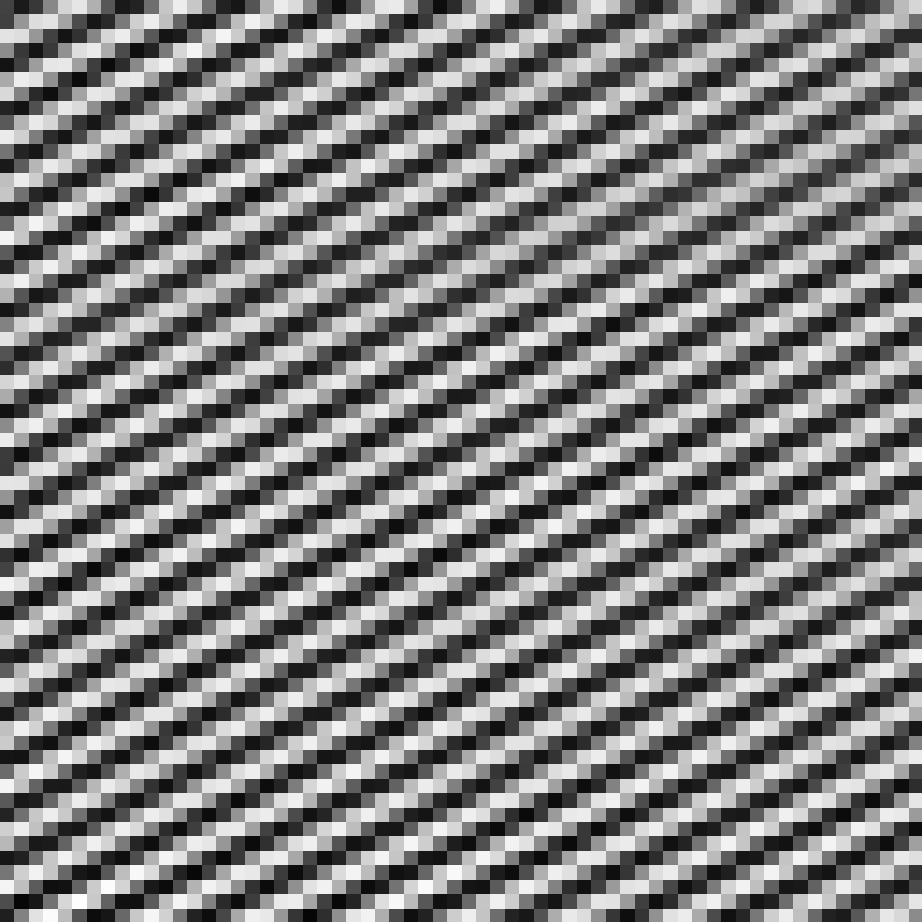}
		\caption{ (101.22 49.29)}
		\label{chutian3}
	\end{subfigure}
	\centering
	\begin{subfigure}{0.325\linewidth}
		\centering
		\includegraphics[width=0.9\linewidth]{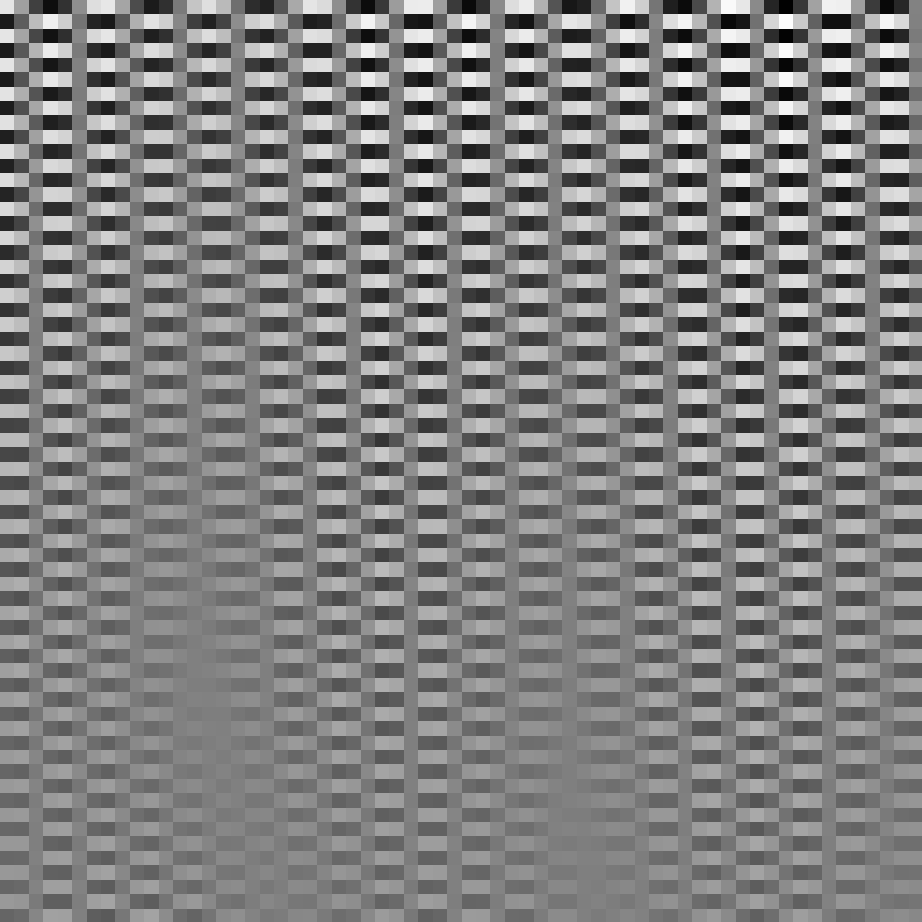}
		\caption{ (73.85 30.77)}
		\label{chutian3}
	\end{subfigure}
	\centering
	\begin{subfigure}{0.325\linewidth}
		\centering
		\includegraphics[width=0.9\linewidth]{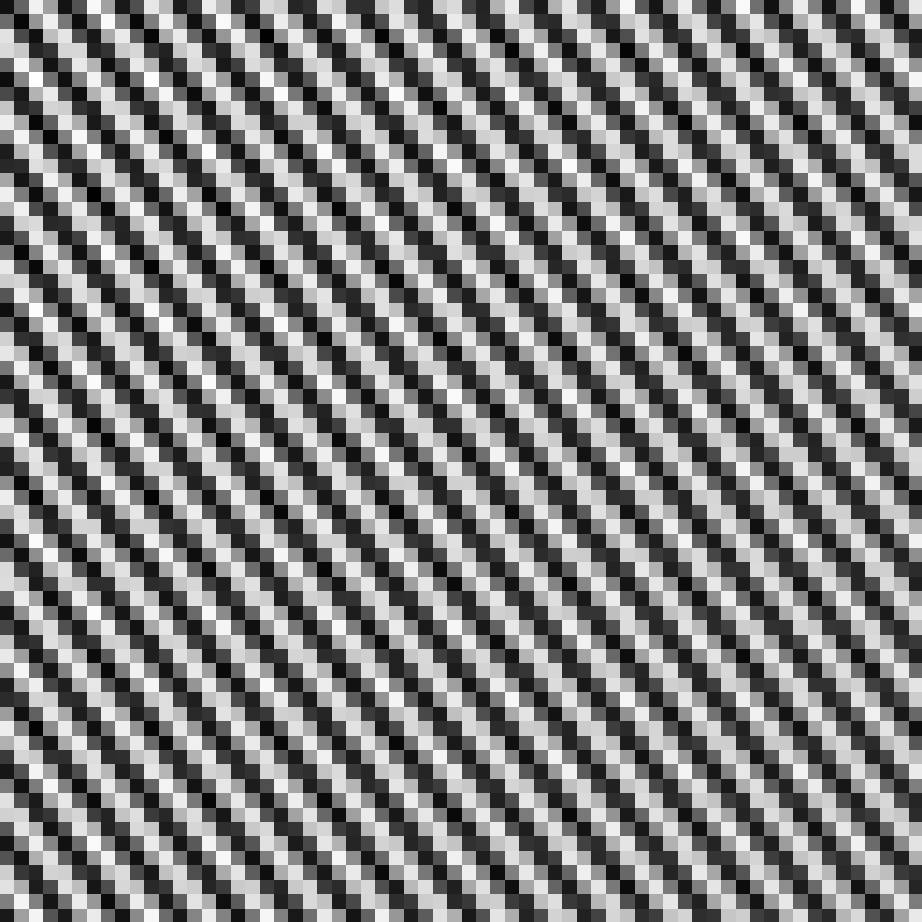}
		\caption{ (144.64 93.75)}
		\label{chutian3}
	\end{subfigure}
	\caption{CFR example figure of randomly selected MTs in the 00743 scenario of WAIR-D}
	\label{fig:CFR_example}
	\vspace{-2mm}
\end{figure}
We commence by depicting the matrix $H$ from Eq. \ref{H} as a  two-channel grayscale image, as illustrated in Fig. \ref{fig:CFR_example}. In this representation, the $x$-axis corresponds to the carrier frequency, while the $y$-axis denotes the antenna pair. A consistent observation across the dataset reveals that the presence of distinct horizontal and vertical translational structures within certain regions in CFR, referred to as PFCFR, serves as a key indicator of CFR data distribution, which is largely shaped by physical phenomena such as refraction, reflection, and diffraction, exhibiting more pronounced similarities in adjacent positions. Consequently, it is a logical inference that CFR sharing similar PFCFR characteristics are likely to be located in geographically close areas. This proximity-based similarity allows for an effective categorization of these CFR according to their respective PFCFR values, providing valuable insights into the spatial dynamics of data distribution influenced by channel properties.

To leverage this insight, we propose a novel two-stage pipeline matched filter outlined in Alg. \ref{alg:PFCFR1} and Alg. \ref{alg:PFCFR2}. Inspired by the target tracking algorithm, the filter is designed firstly to extract PFCFR from all samples as the template and CFR as the source image for CFR categorization. 
To mitigate the risk of coincidental similarities in specific regions between two images, our approach incorporates dual template regions. We define $T_1(i,j,x',y')$ and $T_2(i,j,x',y')$ to represent the pixel values from the upper-left and lower-right corners, respectively, of the $j$-th source image. Here, the notation $(i,x',y')$ specifies the pixel coordinates in the $i$-th template image.
The application of this filter involves systematically shifting the origin of the template image across each point in the source image. The similarity between the template and source images is calculated by aggregating the products of their corresponding pixel values across the entire span of the template. To quantify this similarity, we employ the normalized cross-correlation, denoted as ${\mathbf{E_n(i,j)}}$, given by:

\begin{scriptsize}
\begin{equation}
\hspace{-1.5mm}
    \begin{aligned}
        {\mathbf{E_n(i,j)}}\!&=\! \max\! \frac{\sum_{x',y'}\!{T_n(i,j,x',y')I(j,x\!+\!x',y\!+\!y')}}{\sqrt{\sum_{x',y'}\!{T_n(i,j,x',y')^2} \sum_{x',y'}\!{I(j,x\!+\!x',y\!+\!y')^2}}}{,} 
    \end{aligned}\label{eq:10}\end{equation}
\end{scriptsize}  
where $n={1,2}$, ${\mathbf{E_n(i,j)}}$ is between 0 and 1, and the closer it is to 1, the higher the similarity. Consequently, the process of our proposed two-stage pipeline matched filter can be summarized as follows:

\textbf{Initialization:} 
We set $\tau _c$ to a value greater than $M$, which serves as a threshold for determining whether the $j$-th image is matched. Then, $\tau _{in}$ and $\tau _{out}$ are set as parameters indicative of spatial proximity within and between categories, respectively. Additionally, we use $class\_{num}$ to denote the number of categories.

\textbf{Match Within Categories:}
In this stage, we initiate by organizing the sequence of all images into a list denoted as $path\_List$. Our process starts with the first image in this list, for which we select two distinct templates, represented as $T_n(1,j,x',y')$, where $n=(1,2)$. The subsequent step involves systematically matching the defined template area with corresponding areas in all subsequent images within the $path\_List$, employing Eq. \ref{eq:10} for this purpose. If the match exceeds $\tau _{in}$, the two images are assigned to the same category.
We then move on to the second image. If it hasn't been matched yet, we select two templates, $T_n(2,j,x',y')$, and repeat the matching process. We use the variable $t$ to decide whether a template is matched or not during one matching process.
It's important to note that if the template doesn't find any matching images, the match is sought in the preceding images along the path, starting from the first image. Given that these images have already been categorized, the corresponding category number of the matched image is assigned to the template. We then go to the next image, repeating the above steps. This process continues until each image has been assigned its own category number, denoted as $c_j$. The algorithm is shown as Alg. \ref{alg:PFCFR1}. 

\begin{algorithm}
\caption{Match Within Categories}
\label{alg:PFCFR1}
\begin{algorithmic}[1]
\Require {All CFR images of M users, $\tau _{in}$, $\tau _{c}$, and the size of template $a,b$}
\Ensure $c_j$ for $j = 1,2,...M$
\Initialize{$class\_{num} = 0$ and set $c_j = \tau _c > M$ for $j = 1,2,...M$}
    \For{$i = 1,2,...,M$} \Comment{Match within categories}
    \State $t \gets 0$
    \If{$i=1$ or $c_i = \tau _c$}
      \State Choose $T_n(i,j,x',y')$ according to $a,b$
      \State $c_i \gets class\_{num}$
    \EndIf
        \For{$j = i+1,i+2,...,M$}
            \If{$c_j = \tau _c$}
                \State Compute $E_n(i,j)$
                \If{$E_n(i,j) >= \tau _{in}$}
                    \State $c_j \gets class\_{num}, t \gets 1$
                \EndIf
            \EndIf
        \EndFor
        \If{$t = 0$}
            \For{$k = 1,2,...,i$}
                \State Compute $E_n(i,k)$
                \If{$E_n(i,k) >= \tau _{in}$}
                    \State $t \gets 1$, $c_j \gets c_k$
                    \State Break the loop
                \EndIf
            \EndFor
        \EndIf
        \If{$t = 0$}
            \State $c_j \gets class\_{num}$
        \EndIf
        \State $class\_{num} \gets class\_{num}+1$
    \EndFor
\end{algorithmic}
\end{algorithm}

\textbf{Match Between Categories:}
The objective of this stage is to macroscopically merge similar categories. We utilize Eq. \ref{eq:10} to evaluate the similarity between the templates derived from the first-stage classification. When the similarity between any two templates exceeds $\tau _{out}$, we amalgamate all CFR images associated with these templates into a single category. This merging process continues until there is no further change in the number of final classifications. The comprehensive algorithm for this stage is methodically outlined in Alg. \ref{alg:PFCFR2}.

\begin{algorithm}
\caption{Match Between Categories}
\label{alg:PFCFR2}
\begin{algorithmic}[1]
\Require {All templates and respective representative CFR images, $\tau _{out}$, $c_j$ and the size of template $a,b$}
\Ensure $c_j$ for $j = 1,2,...M$
    \State {Arrange $c_j$ in ascending order and update the index of the arranged $c_j$ to the new $c_j$ for $j = 1,2,...M$}
    \For{$i = 1,2,...,max(c_j)$} \Comment{Match between categories}
        \For{$j = 1,2,...,max(c_j)$}
            \If{$j \neq i$}
                \State Choose $T_n(i,j,x',y')$ according to $a,b$
                \State Compute $E_n(i,j)$
                \If{$E_n(i,k) >= \tau _{out}$}
                    \State Merge $c_i$ and $c_j$ into the same category
                \EndIf
            \EndIf
        \EndFor
    \EndFor
    \State {Arrange $c_j$ in ascending order and update the index of the arranged $c_j$ to the new $c_j$ for $j = 1,2,...M$}
\end{algorithmic}
\end{algorithm}

\subsubsection{ADCAM}
We commence by depicting the matrix ${A}_m$ from Eq. \ref{A} as a two-channel grayscale image.
To assist in the pre-classification of ADCAM, we employ the iterative centroid-based clustering method to partition information such as the AOD, AOA, gains, and pathloss of each path into $K$ clusters based on their distances from $K$ class centers. 
One major challenge is to determine the appropriate value of $K$. Usually, it is chosen based on industry experience and may not necessarily reflect the true number of clusters present in the data. 
To address this challenge, we combine the Silhouette coefficients (SC) \cite{aranganayagi2007clustering} and Calinski-Harabasz (CH) methods \cite{maulik2002performance} to combine both the direct distances and covariance. This helps avoid scenarios where the graph of the clustering quality measure exhibits a smooth pattern (e.g., horizontal or continuously ascending/descending), making it difficult to ascertain the ideal values \cite{jain2010data}.

\textbf{SC method:} We calculate the average silhouette for each candidate value. The value that corresponds to the highest average silhouette score, denoted as $\mathbf{S_{SC}(i)}$, is considered the optimal number of clusters for the unsupervised learning algorithm. The silhouette score for a single data point, $S_{SC}(i)$, is computed as:
\begin{equation*}
    {\mathbf{S_{SC}(i)}}=\frac{b(i)-a(i)}{max\{a(i),b(i)\}} 
,\tag{11}\end{equation*}
where ${{a(i)}}=\frac{1}{n-1}\sum_{j\neq i}^n distance(k(i),k(j))$ gives a measure of how well assigned the $i$-th data point is to it’s cluster, and ${{b(i)}}=\mathop{\min}_{j\neq i} \frac{1}{n}\sum^n distance(k(i),k(j))$ gives the average dissimilarity to the closest cluster which is not it’s cluster, ${{k(i)}}=[AOD_i, AOA_i, gains_i, pathloss_i]$. 

\textbf{CH method:} The CH index serves as a measure of how similar an object is to its own cluster (cohesion) relative to other clusters (separation). Cohesion is estimated based on the distances between data points within a cluster and their cluster centroid, while the separation is computed based on the distances between cluster centroids and the global centroid.

Unlike the SC method, which focuses on direct distances, the CH method considers the covariance for clustering quality. The CH index, denoted as $\mathbf{S_{CH}(k)}$, is calculated as follows:
\begin{equation*}
    {\mathbf{S_{CH}(k)}}=\frac{tr(B_k)(q-k)}{tr(W_k)(k-1)} 
,\tag{12}\end{equation*}
where $B_k$/$W_k$ is the covariance matrix between/within categories with $k$ categories, and $q$ is the number of samples in the training set.

\subsubsection{Data Fusing and Cleansing}
To ensure high-quality input data for our model, we combine the classification results obtained from both the PFCFR and ADCAM. This integration step helps us create a comprehensive and unified classification scheme. Here's how it works:

\textbf{Combining Classifications:}
our initial step involves utilizing the two classification regions to refine the final classification categories. Subsequently, we apply data cleansing techniques to remove any anomalous fingerprints that might skew the analysis.
In the phase of combining classifications, we adopt a systematic approach. For instance, if there are only Class 0 and Class 5 ADCAM images found within the Class 0 CFR category, we assign values of [0, 0] and [0, 5] to the new categories as Class 0 and 1, respectively. This reclassification process is methodically applied across all categories.

\textbf{Data Cleansing:}
As previously mentioned, there may be special fingerprints in the scene that pose challenges that traditional methods struggle to handle. 
After our pre-processing, we can apply data cleansing techniques to address this issue. 
We set a number to eliminate categories with a sample size of less than or equal to the number. The rationale behind this decision is that such a small sample size in a category typically indicates an aberration, possibly due to multi-path features that significantly diverges from neighboring points. This could be attributed to erroneous recordings or the point being situated in a distinctly unique location. 

\subsection{Network Training}
After classifying the samples into different regions as shown in Fig. \ref{fig:visulizaion1}, we proceed with the network training process. This process consists of feature extraction, feature fusion, and filter regression, as illustrated in Fig. \ref{fig:framework}. The objective is to enable fingerprint localization, expressed as:
\begin{equation*}
    {\hat{\bf {p}}_{i,j}}={\mathbf{\mathcal G_{3,i}}}(\mathbf{\mathcal P}(\mathbf{{\mathcal G_{1}}}(H_{i,j}),{\mathbf{\mathcal G_{2}}(A_{i,j})}))
,\tag{9}\end{equation*}
where ${\mathbf{\mathcal G_{1}}}$ and ${\mathbf{\mathcal G_{2}}}$ represent the mappings built by ResNet18 \cite{he2016deep} from CFR or ADCAM images to a vector with a flattened length of $M_H$ or $M_A$ respectively. $\mathbf{\mathcal P}$ is a function that merges these two sets of features along the same dimensions. 
Finally, $\mathbf{\mathcal G_{3,i}}$ is a segment-specific linear classifier mechanism in which the combined features are divided into different linear classifiers for regression operations. The variables $\sum_i i = I$ denote the number of final classifications, and each linear classifier corresponds to a specific region denoted as $i$. $\sum_i \sum_j i\times j = M$, and ${i,j}$ represents the $j$-th sample in the $i$-th class of the fingerprint after the heterogeneity matching. 

The loss function used for training is the mean-squared error (MSE) between the true positional coordinates $p$ and the predicted positional coordinates $\hat{p}$, calculated as:
\begin{equation*} \text {MSE Loss}=\frac {1}{n}\sum _{i=1}^{n}\left \|{\hat { \boldsymbol {p}}_{i}- \boldsymbol {p}_{i}}\right \|^{2},\tag{10}\end{equation*}
Here, $n = 16$ indicates the mini-batch size.

\section{Numerical Results}\label{sec:intro}

In this section, we utilize dense MTs SCMA-OFDM scenarios in WAIR-D and DeepMIMO datasets to validate the effectiveness, interpretability, and scalability of the AMDNLoc framework across various scenarios. 
Our approach was trained for 150 epochs with the Adam optimizer with a learning rate of $3\times 10^{-3}$ in the Nvidia 3090. 
Weight initialization is performed using Xavier initialization, and the Adam optimizer is utilized. The initial learning rate is set to 0.003.
Network architectures and hyperparameters were chosen via cross-validation for consistency across experiments unless stated otherwise.

\subsection{Datasets}
WAIR-D is a mmWave MIMO dataset for researchers to study and evaluate AI algorithms for wireless systems. It uses the ray-tracing method to approximate the Maxwell equations and describes the propagating field as a set of propagating rays, reflecting, diffracting, and scattering over environment elements to obtain the accurate characterization and simulation of electromagnetic propagation \cite{choi2023withray}. 
DeepMIMO is a widely used dataset in the industry, known for its general applicability and robustness. It has been the foundation for numerous research studies due to its comprehensive modeling approach, which combines scene construction with ray-tracing methods. This combination provides highly accurate and reliable data, making DeepMIMO a benchmark in the field.
In comparison to the widely-used DeepMIMO dataset, WAIR-D offers a more diverse range of practical propagation environments. This diversity makes it particularly suitable for training robust generalization capabilities in AI models \cite{fenghao1}.

The parameters employed for generating these datasets are detailed in Tab. \ref{tab:dataset}. For the purpose of our experiments, we assumed that all MTs are single-antenna user devices (UEs).
We believe the two chosen datasets already provide a comprehensive evaluation of our approach's performance and effectiveness.
This comprehensive approach allows us to assess the adaptability of AMDNLoc to different environmental conditions and parameters, demonstrating its potential for wide applications in wireless communications research.
\begin{table}[h!]
\normalsize
\centering
 \vspace{-2mm}
 \caption{Datasets parameters.}
 \begin{tabular}{c c | c c} 
 \toprule
 Parameters & Value & Parameters & Value\\ [0.2ex] 
 \midrule
 Carrier frequency & 60 GHz & BS antenna & [1, 64, 1]\\ 
 Bandwidth & 0.05 GHz & UE antenna & [1, 1, 1] \\
 Sub-carriers & 64 & Size of template & [8, 16] \\
 \bottomrule
 \end{tabular}
 \label{tab:dataset}
 \vspace{-4mm}
\end{table}

\subsection{Visualization of Classification Regions}
\begin{figure*}[htbp]
	\centering
	\begin{subfigure}{0.325\linewidth}
		\centering
		\includegraphics[width=1\linewidth]{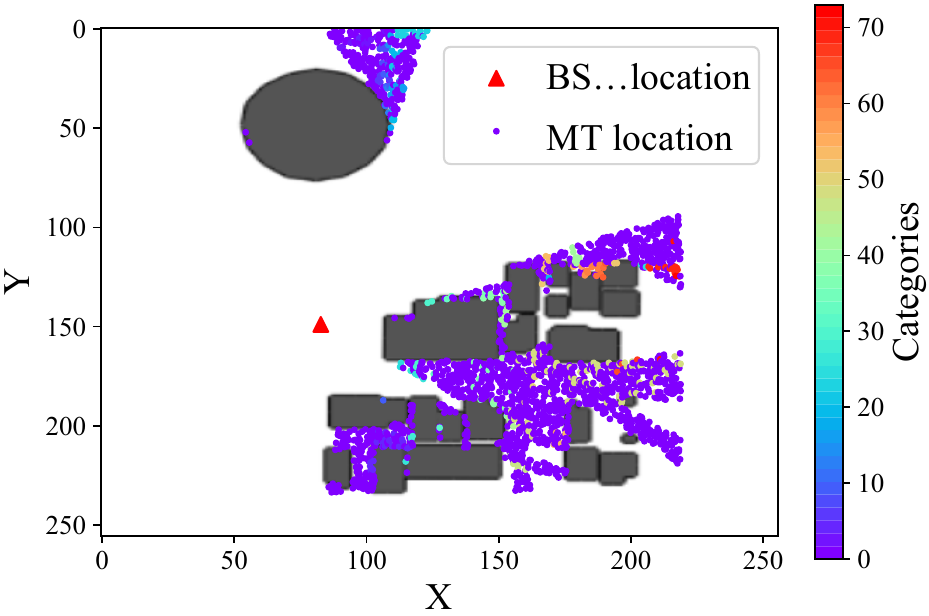}
		\caption{frequency domain}
		\label{fig:visulizaion1.1}
	\end{subfigure}
	\centering
	\begin{subfigure}{0.325\linewidth}
		\centering
		\includegraphics[width=1\linewidth]{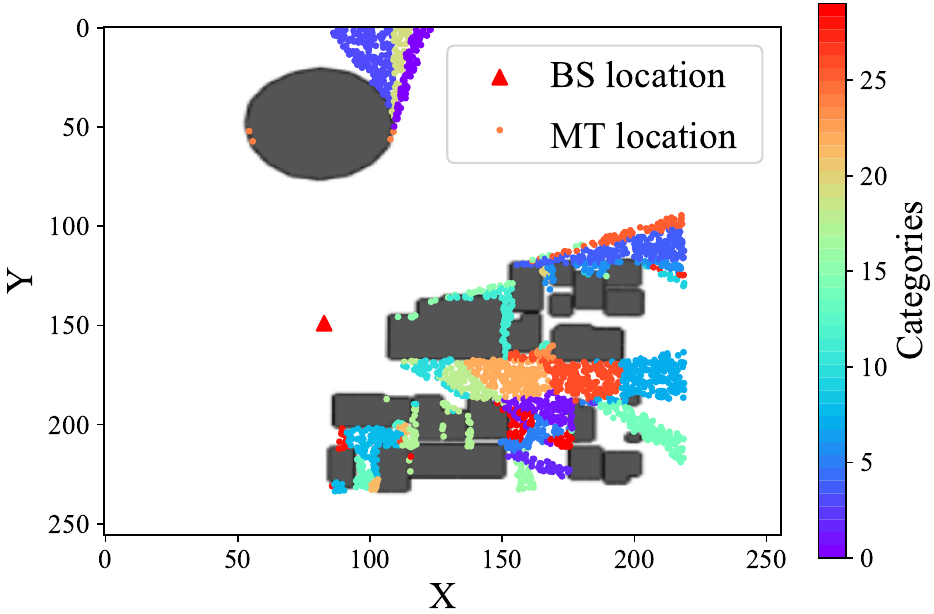}
		\caption{power, angle and delay domain}
		\label{fig:visulizaion1.2}
	\end{subfigure}
	\centering
	\begin{subfigure}{0.325\linewidth}
		\centering
		\includegraphics[width=1\linewidth]{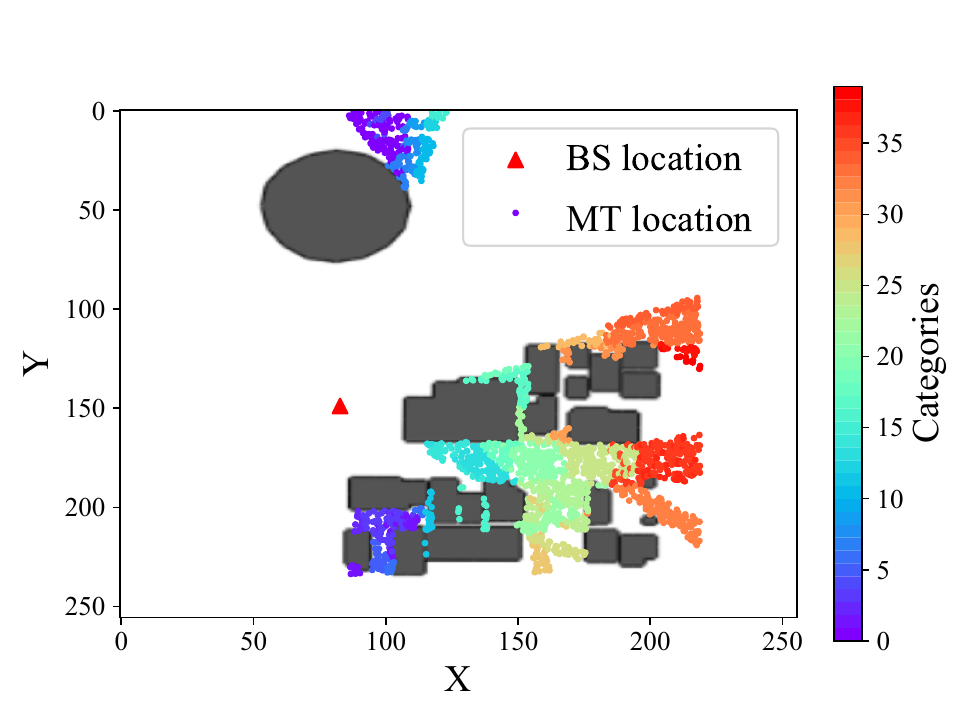}
		\caption{combining domain}
		\label{fig:visulizaion1.3}
	\end{subfigure}
	\caption{NLOS pre-classification in multi domains when $\tau _{in}=0.99, \tau _{out}=0.99$}
	\vspace{-2mm}
	\label{fig:visulizaion1}
\end{figure*}
\begin{figure*}[htbp]
	\centering
	\begin{subfigure}{0.325\linewidth}
		\centering
		\includegraphics[width=1\linewidth]{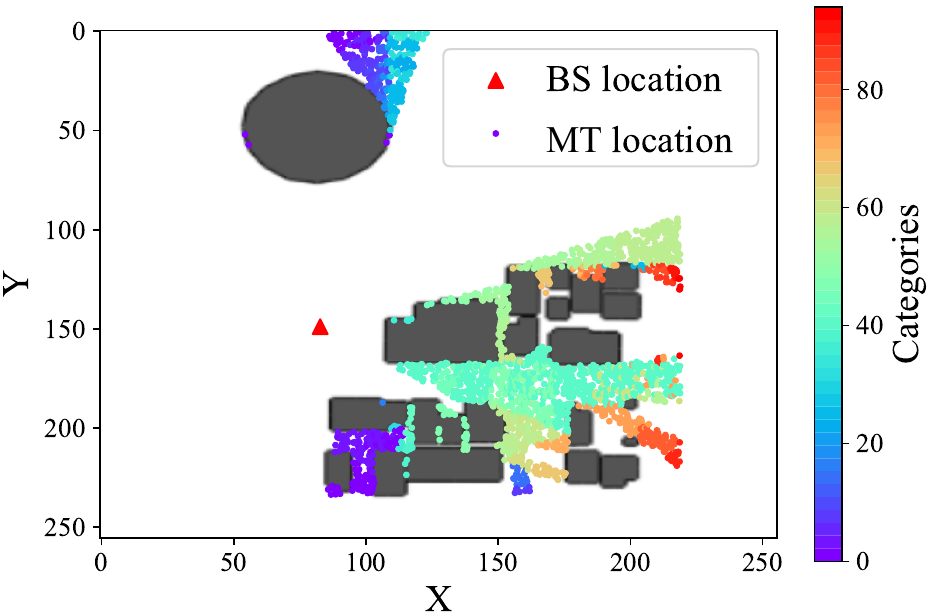}
		\caption{$\tau _{in}=0.98, \tau _{out}=0.99$}
		\label{chutian3}
	\end{subfigure}
		\centering
	\begin{subfigure}{0.325\linewidth}
		\centering
		\includegraphics[width=1\linewidth]{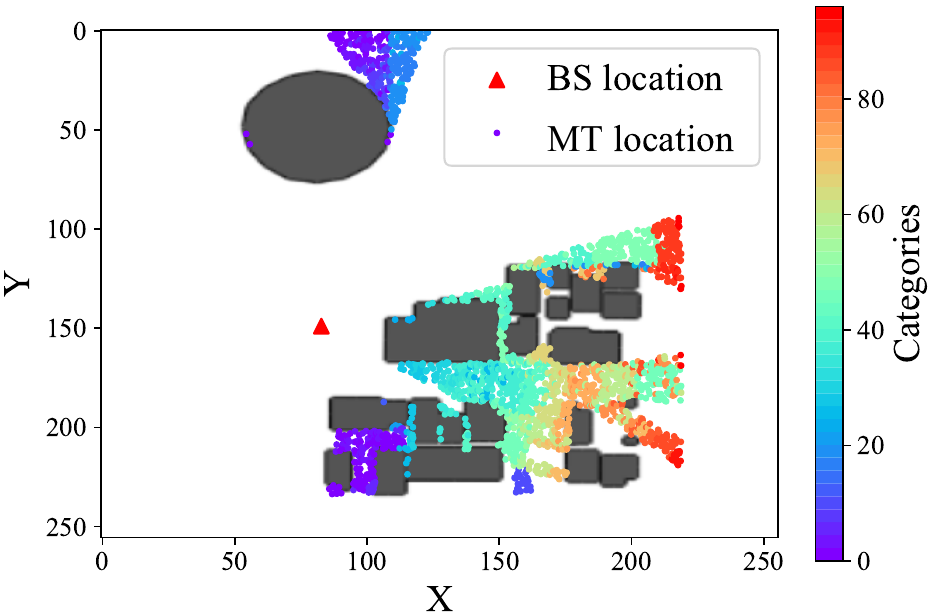}
		\caption{$\tau _{in}=0.9, \tau _{out}=0.99$}
		\label{chutian3}
	\end{subfigure}	\centering
	\begin{subfigure}{0.325\linewidth}
		\centering
		\includegraphics[width=1\linewidth]{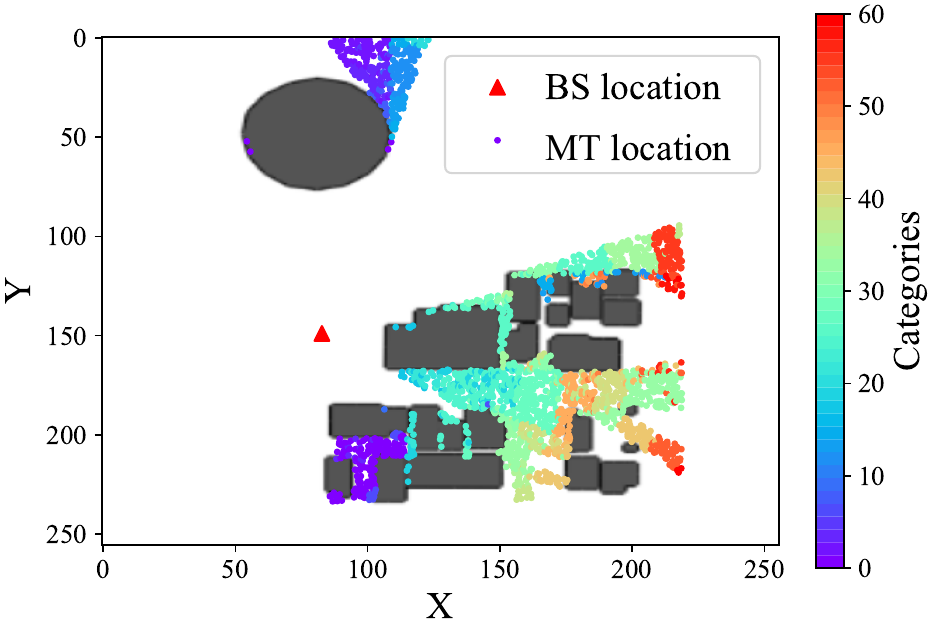}
		\caption{$\tau _{in}=0.8, \tau _{out}=0.99$}
		\label{chutian3}
	\end{subfigure}	\centering
	\caption{NLOS pre-classification when change $\tau _{in}$ in frequency domain}
	\vspace{-2mm}
	\label{fig:visulizaion2}
\end{figure*}
\begin{figure*}[htbp]
	\begin{subfigure}{0.325\linewidth}
		\centering
		\includegraphics[width=1\linewidth]{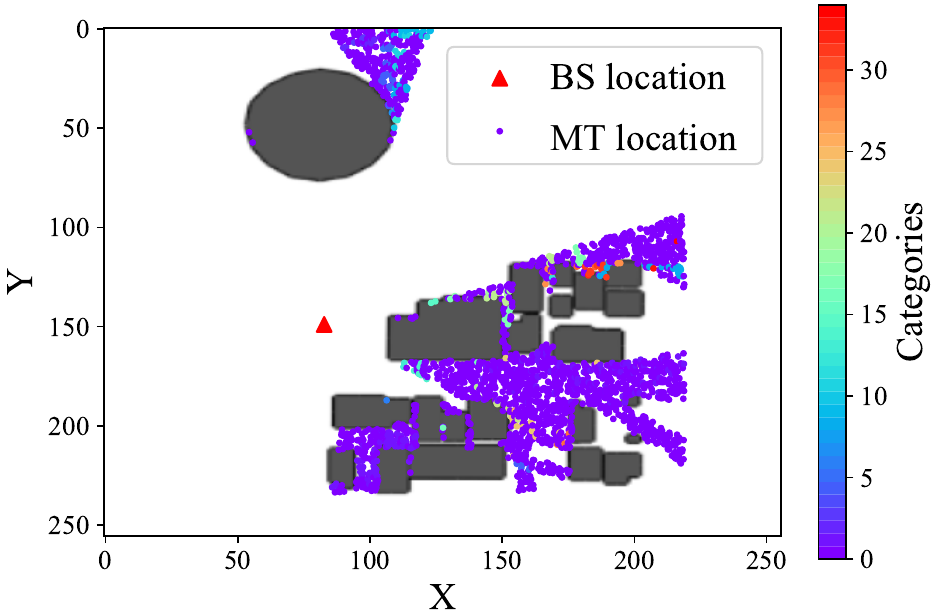}
		\caption{$\tau _{in}=0.99, \tau _{out}=0.98$}
		\label{chutian3}
	\end{subfigure}	\centering
	\begin{subfigure}{0.325\linewidth}
		\centering
		\includegraphics[width=1\linewidth]{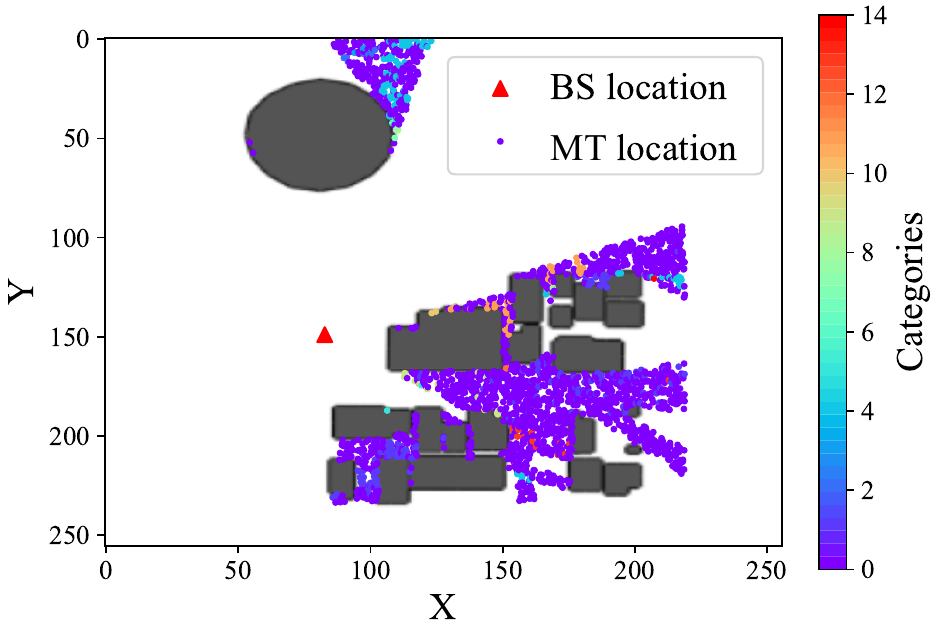}
		\caption{$\tau _{in}=0.99, \tau _{out}=0.95$}
		\label{chutian3}
	\end{subfigure}	\centering
	\begin{subfigure}{0.325\linewidth}
		\centering
		\includegraphics[width=1\linewidth]{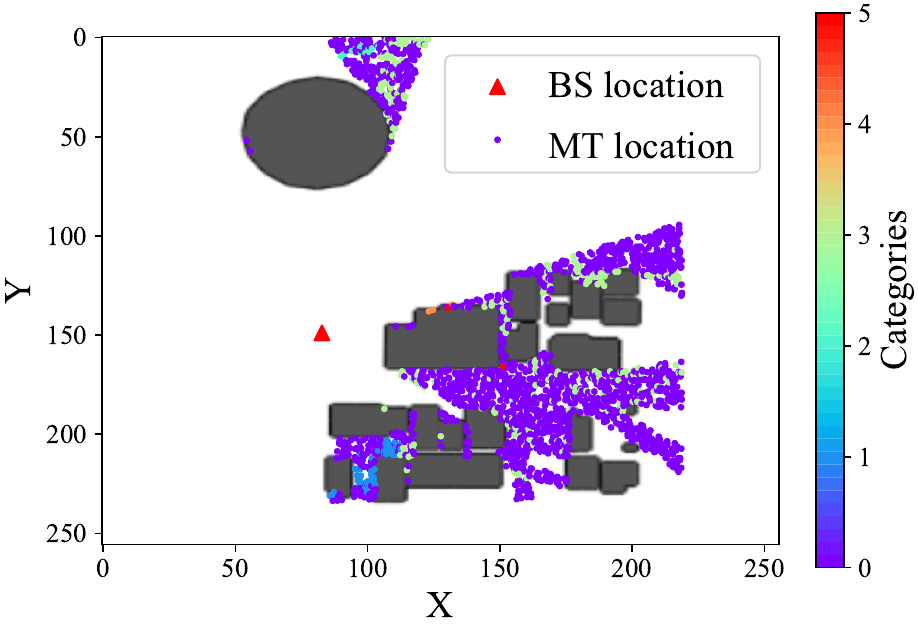}
		\caption{$\tau _{in}=0.99, \tau _{out}=0.9$}
		\label{chutian3}
	\end{subfigure}
	\caption{NLOS pre-classification when change $\tau _{out}$ in frequency domain}
	\label{fig:visulizaion3}
	\vspace{-3mm}
\end{figure*}
\begin{table}[h!]
\normalsize
\centering
 \vspace{-2mm}
 \caption{Neural networks parameters.}
 \begin{tabular}{c c | c c} 
 \toprule
 Parameters & Value & Parameters & Value\\ [0.2ex] 
 \midrule
 Batchsize & 16 & Loss function & MSE\\ 
 Learning rate & 0.001 & Activation function & ReLU \\
 Epochs & 50 & Optimizer & Adam  \\
 \bottomrule
 \end{tabular}
 \label{tab:neuralnet}
 \vspace{-1mm}
\end{table}
Firstly, we choose the resolution of 16x16 pixels of $T_1$ and $T_2$. Then, we know that the division of NLOS regions within AMDNLoc is primarily guided by the termination thresholds, $\tau _{in}$ and $\tau _{out}$, in the PFCFR matched filter. 
Initially, we visualize the pre-classification regions across multiple domains with $\tau _{in}=\tau _{out}=0.99$, as depicted in Fig. \ref{fig:visulizaion1}. Notably, Fig. \ref{fig:visulizaion1.3} demonstrates more flexible and reasonable boundaries, attributable to the deep fusion of features in combining domains.

Subsequently, to assess the impact of varying termination thresholds on the localization, we alter $\tau _{in}$ and $\tau _{out}$ independently, as illustrated in Fig. \ref{fig:visulizaion2} and Fig. \ref{fig:visulizaion3}. We observe that changes in $\tau _{out}$ significantly affect the number of categories more than that in $\tau _{in}$. This is because $\tau _{out}$ constrains the inter-category similarity, and a lower value tends to merge more categories. Hence, it acts as a coarse tuning threshold and should be set to a higher value. Conversely, $\tau _{in}$ governs the intra-category similarity, influencing the classification of fingerprints into categories. Adjusting $\tau _{in}$ therefore has a fine tuning effect on the region division.

To determine optimal values for $\tau _{out}$ and $\tau _{in}$, we compute the MSE by varying $\tau _{in}$ while keeping $\tau _{out}$ constant at 0.99, as shown in Tab. \ref{tab:tau}. Additionally, we vary the number of data cleansing samples to 2 or 10. We find that decreasing $\tau _{in}$ reduces MSE to a certain extent. Furthermore, removing more special categories leads to smaller scene coverage but higher localization accuracy. Although a trade-off exists between coverage area and MSE, opting for a lower coverage rate can enhance localization accuracy. This approach aligns with our expectations, considering that if a MT encounters a special sample, it can easily move a short distance to achieve more accurate position estimation.

In conclusion, our pre-classification method is highly effective and reliable. It divides a large outdoor area of $250\times250$ meters, with numerous buildings, into small regions with irregular boundaries. Each region exhibits similar fingerprint distribution characteristics, and there is a clear distinction between regions, significantly mitigating the issue of fingerprint distribution heterogeneity. 
These characteristics make our approach applicable in various contexts such as channel estimation \cite{9526282, xiao2022c, liu2022deep}, channel extrapolation \cite{zhang2021deep}, CSI design for sensing \cite{gan2021ris, tong2021joint}, and beamforming \cite{10531779, huang2021multi}.
\begin{table}[h!]
\normalsize
\centering
 \caption{The MSE of changing the $\tau _{in}$ when $\tau _{out}=0.99$.}
 \begin{tabular}{c c c c} 
 \toprule
 $\tau _{in}$ & \makecell[c]{Data Cleansing \\ Sample Number} & \makecell[c]{Region \\ Covering Rate} & MSE(m) \\  [0.2ex] \midrule

 \multirow{2}{*}{0.99} & 2 & 0.984 & 2.54 \\  & 10 & 0.918 & 2.23 \\ \hline
 \multirow{2}{*}{0.98} & 2 & 0.964 & 2.66 \\  & 10 & 0.823 & 2.28 \\ \hline
 \multirow{2}{*}{0.9} & 2 & 0.967 & 2.62 \\  & 10 & 0.782 & 2.57 \\ \hline
 \multirow{2}{*}{0.8} & 2 & 0.976 & 2.36 \\  & 10 & 0.824 & 2.17 \\
 \bottomrule
 \end{tabular}
 \label{tab:tau}
 \vspace{-3mm}
\end{table}

\subsection{Effectiveness of Classification Methods}
\subsubsection{Comparison of Multiple Processing Results (MPR)}
The neural networks parameters are shown in the Tab. \ref{tab:neuralnet}. In this part, we compare our proposed algorithm with several baseline algorithms for NLOS environments using a ULA. To facilitate a fair comparison, we uniformly employ ResNet18 as the backbone for feature extraction in all methods. The feature extractors for CFR and ADCAM in different comparison methods share the same structure to maintain consistency. The structures of all AI-based baselines have been carefully optimized through cross validation to ensure fairness in comparison. The baseline algorithms include:
\begin{enumerate}
    \item Res\_CFR/Res\_ADCAM/Res\_CFRADCAM: These methods utilize either CFR, ADCAM, or a combination of both as input for the feature extractor before feeding the features into a linear classifier to predict coordinates.
    \item Res\_multi\_CFRADCAM/Res\_multi\_CFRperfectADCAM: Building on Res\_CFRADCAM, these baselines input the obtained features into multiple linear classifiers for the coordinate prediction. Specifically, Res\_multi\_CFRperfectADCAM replaces ADCAM with additional inputs such as AOA, AOD, distance, and gain from the first arrival path.
\end{enumerate}
\begin{figure}\vspace{-2mm}
	\begin{center}
		\centerline{\includegraphics[width=0.45\textwidth]{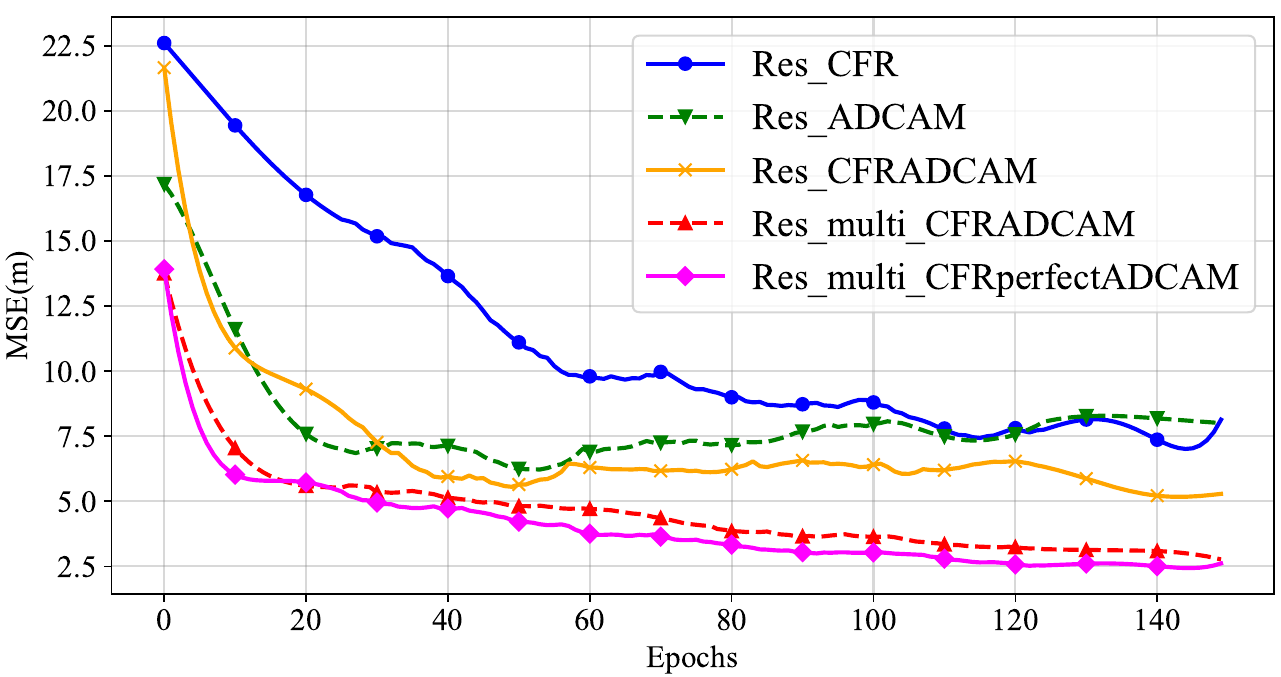}}  \vspace{-0mm}
		\caption{MSE with epochs of MPR.}
		\label{fig:multiplemseepoch} \vspace{-8mm}
	\end{center}
\end{figure}
\begin{figure}\vspace{-0mm}
	\begin{center}
		\centerline{\includegraphics[width=0.45\textwidth]{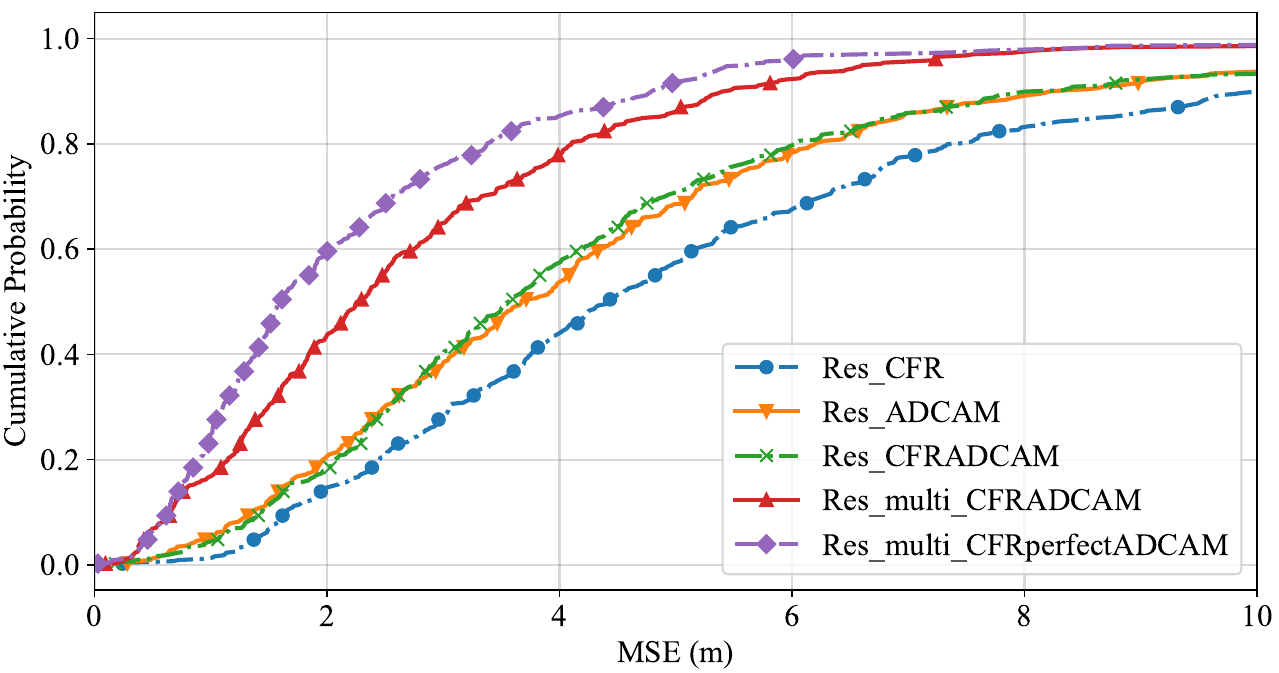}}  \vspace{-0mm}
		\caption{CDF curves of MPR.}
		\label{fig:multiplecdf} \vspace{-9mm}
	\end{center}
\end{figure}

Fig. \ref{fig:multiplemseepoch} provides convincing evidence of our methodology's effectiveness. Initially, we observe that the MSE for Res\_ADCAM decreases rapidly with an increase in epochs but eventually stabilizes near the higher MSE of Res\_CFR. This pattern emerges because, while ADCAM is sparser than CFR, it does not contain as much information, so using complex network structures is more prone to overfitting.
Then, we notice a significant improvement when combining CFR and ADCAM as input data. By preprocessing and fusing these inputs, and employing a segment-specific linear classification mechanism, our results show a remarkable improvement, reducing the error from 9 meters to 2.54 meters in just 150 epochs. This improvement is attributed to AMDNLoc's focus on the variations between locations and fingerprints with similar data distributions, allowing for a more accurate reflection of the true nature of the localization-fingerprint relationship.

Moreover, Fig. \ref{fig:multiplecdf} underscores this advancement, revealing that over 60\% of positioning errors in our model fall within a 2-meter range. In contrast, the baseline models only demonstrate a 20\% probability of achieving such accuracy. These findings unequivocally affirm the superiority of our integrated approach in enhancing the accuracy and reliability of positioning in multiple MTs SCMA-OFDM systems.

\subsubsection{Comparison of Pre-processing on CFR or ADCAM}
\begin{figure}\vspace{-2mm}
	\begin{center}
		\centerline{\includegraphics[width=0.43\textwidth]{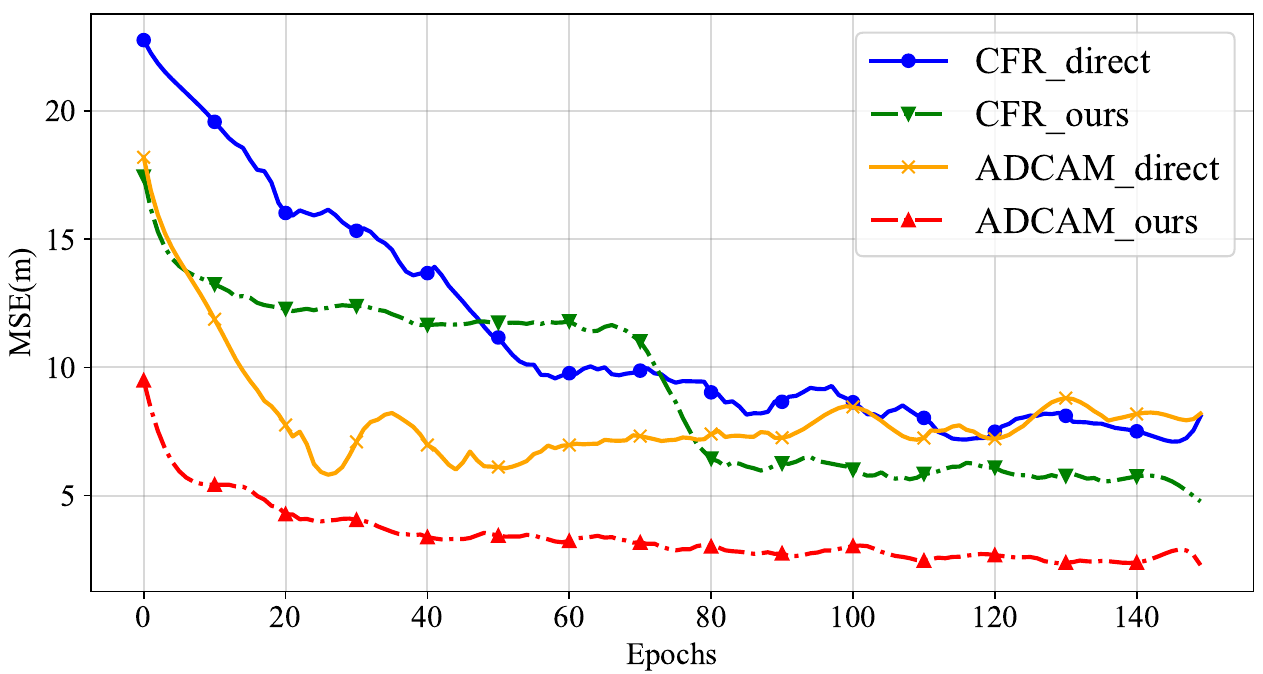}}  \vspace{-0mm}
		\caption{MSE with epochs on CFR and ADCAM.}
		\label{fig:multiCFRADP} \vspace{-8mm}
	\end{center}
\end{figure}

In this part, we evaluate the performance of CFR and ADCAM as inputs for feature extraction and localization, both individually and in conjunction with our pre-classification technique as illustrated in Fig. \ref{fig:multiCFRADP}. The approaches are as follows:
\begin{enumerate}
    \item CFR/ADCAM\_direct: This method involves using either CFR or ADCAM as the input for the feature extractor. The extracted features are then processed through a linear classifier to predict coordinates.
    \item CFR/ADCAM\_ours: This approach employ multiple linear classifiers for coordinate prediction building upon CFR/ADCAM\_direct. The number of classifiers corresponds to the number of pre-classifications in the frequency or power, angle, and delay domains.
\end{enumerate}

The results, as depicted in Fig. \ref{fig:multiCFRADP}, reveal some key insights. Firstly, we observe that using ADCAM alone as a fingerprint for coordinate prediction yields better results than using CFR alone, which aligns with the existing results in the \cite{sun2019fingerprint}. Secondly, upon applying our PFCFR matched filter, the performance of CFR significantly improves, eventually surpassing that of ADCAM as the iterations progress. Similarly, a substantial enhancement is noted in the use of ADCAM after iterative clustering. These findings indicate that our pre-classification method, combined with an array of linear classifiers, is highly effective and valuable. This holds true not only for the combined use of CFR and ADCAM, but also when each is used independently.

\subsection{Impact of System Parameters}
In this subsection, the impact of various system parameters are investigated to validate the superiority and universality of the proposed approach.
\subsubsection{Antenna Array}
\begin{figure}\vspace{-0mm}
	\begin{center}
		\centerline{\includegraphics[width=0.45\textwidth]{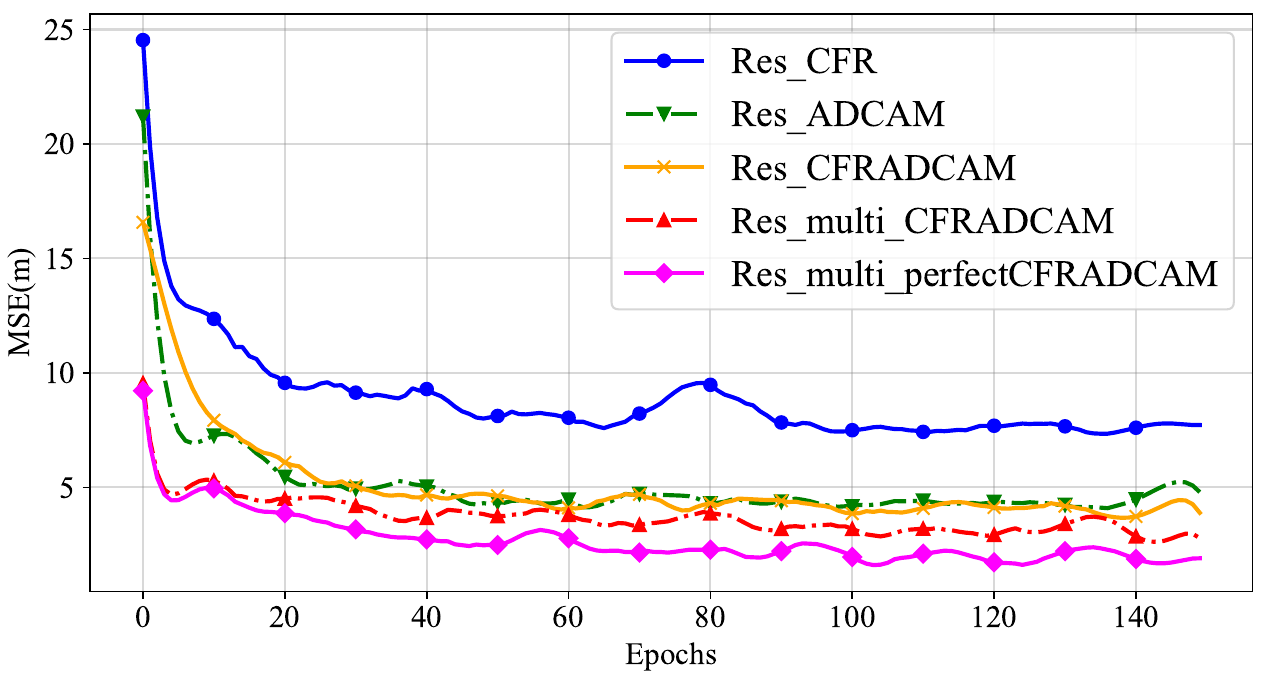}}  \vspace{-1mm}
		\caption{MSE with epochs of MPR with UPA array.}
		\label{fig:multiplemseepoch_188} \vspace{-8mm}
	\end{center}
\end{figure}
\begin{figure}\vspace{-2mm}
	\begin{center}
		\centerline{\includegraphics[width=0.45\textwidth]{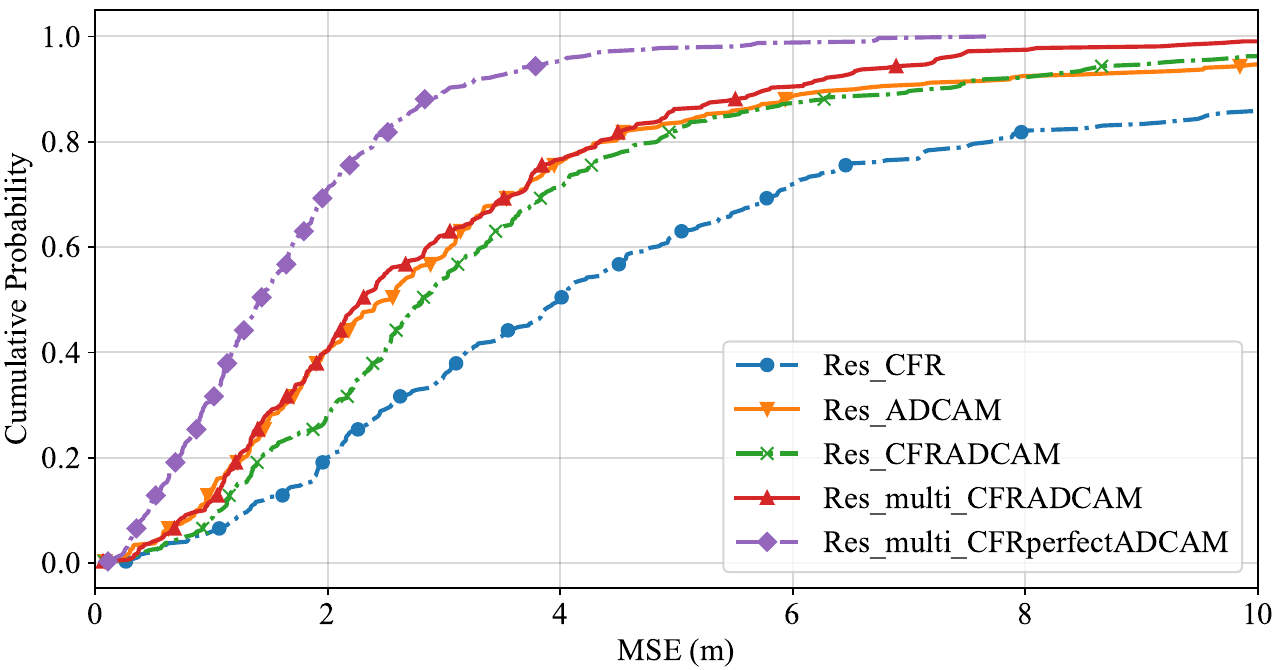}}  \vspace{-0mm}
		\caption{CDF curves of MPR with UPA array.}
		\label{fig:multiplecdf_188} \vspace{-9mm}
	\end{center}
\end{figure}
We alter the transmitting antennas from ULA with a configuration of [1,64,1] to UPA of [1,8,8]. The outcomes of this change are depicted in Fig. \ref{fig:multiplemseepoch_188} and Fig. \ref{fig:multiplecdf_188}.
Upon comparison with the results from the ULA array shown in Fig. \ref{fig:multiplecdf} and Fig. \ref{fig:multiCFRADP}, we observe that the trend and relative positioning of each curve do not exhibit significant differences. 
However, it is noteworthy that, despite the total number of transmitting antennas remains the same, the planar array configuration yields better results in terms of MPR. With the implementation of AMDNLoc, the positioning error decreases from 6.95 meters with CFR to 1.46 meters, which constitutes a remarkable 79\% enhancement in the localization accuracy for outdoor multi-points NLOS scenarios.

\subsubsection{Signal to Noise Ratio (SNR)}
\begin{figure}\vspace{-2mm}
	\begin{center}
		\centerline{\includegraphics[width=0.45\textwidth]{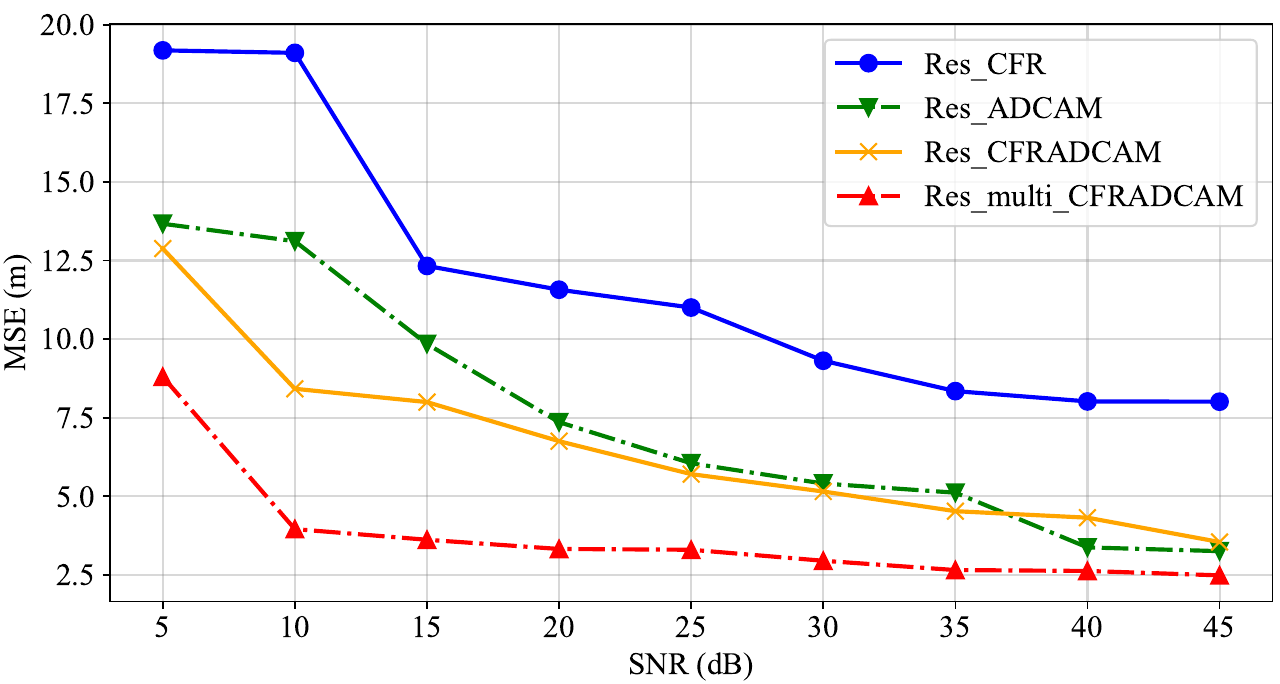}}  \vspace{-0mm}
		\caption{MSE with SNR of MPR.}
		\label{fig:multiSNR} \vspace{-8mm}
	\end{center}
\end{figure}

To analyze the impact of SNR on the localization accuracy, we first normalize the energy of the transmitted signal and calculate the average energy of the channel. Then, based on the SNR, we determine the variance of the noise and generated a complex Gaussian matrix. This matrix is incorporated into the fingerprint to observe the variation in the optimal MSE with varying SNR levels, as illustrated in Fig. \ref{fig:multiSNR}.

Our findings show that although the accuracy of localization of MPR generally improves with an increase in SNR, the variation in the results of AMDNLoc with respect to SNR is comparatively minimal. Notably, in 10-15dB, AMDNLoc demonstrates localization estimation results that are comparable to those achieved at higher SNR levels and significantly better than those obtained without using AMDNLoc. This performance indicates that AMDNLoc is more robust and better suited to the diverse conditions of real-world scenarios, especially where SNR may vary.

\subsubsection{Maxpathnum}
\begin{figure}\vspace{-0mm}
	\begin{center}
		\centerline{\includegraphics[width=0.45\textwidth]{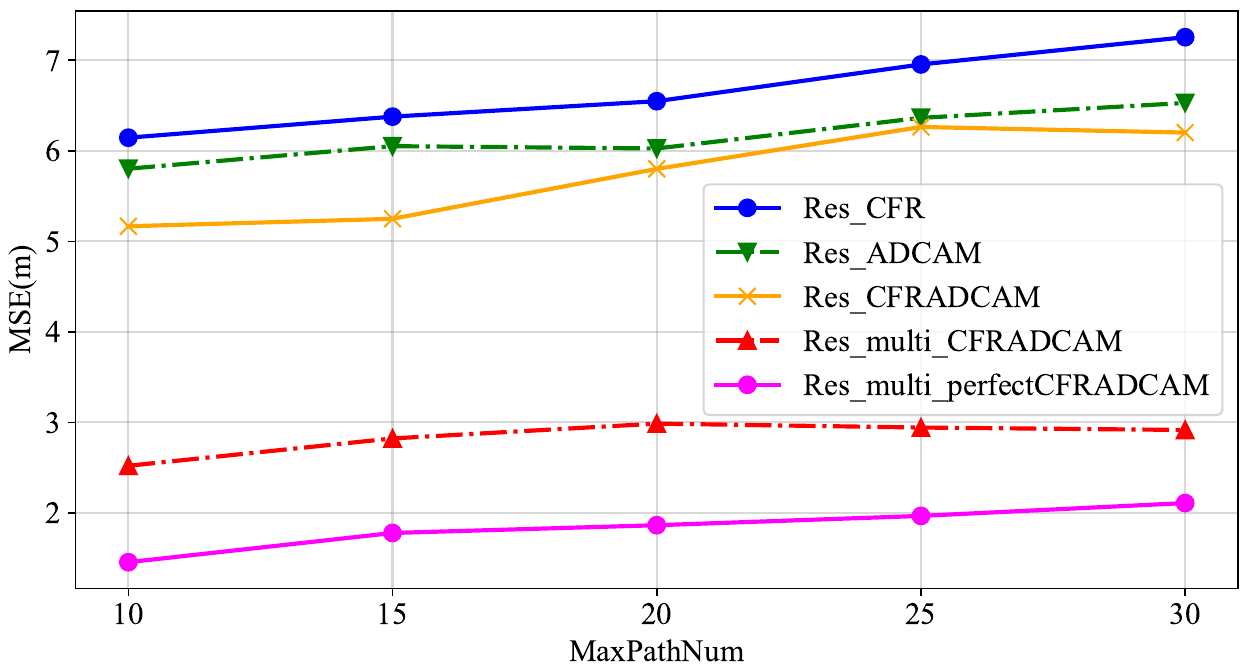}}  \vspace{-0mm}
		\caption{MSE with Maxpathnum of MPR.}
		\label{fig:mutlimaxpathnum} \vspace{-8mm}
	\end{center}
\end{figure}

Maxpathnum, representing the maximum number of paths between a transmitter and a receiver, is a crucial parameter in ray-tracing software. A higher Maxpathnum value implies more scattering and refraction, leading to increased complexity in the multi-paths effect. As shown in Fig. \ref{fig:mutlimaxpathnum}, when Maxpathnum is incrementally increased from 10 to 30, there is a noticeable degradation in the optimal MSE of MPR, reflecting the intensified multi-paths effects.

In contrast, the localization error of AMDNLoc in relation to Maxpathnum remains smaller compared to other methods. This can be attributed to the deep fusion of features across multiple domains within AMDNLoc. Such fusion ensures that variations in the fingerprint dataset do not significantly impact the overall classification regions. Consequently, our framework exhibits greater stability and adaptability, particularly in static environments with dynamic changes. This ability to maintain accuracy despite increasing Maxpathnum underscores the robustness of AMDNLoc in complex multi-paths scenarios.

\subsection{Generalization Ability}
\subsubsection{Multiple Scenarios and Datasets}
We conduct a comprehensive validation of the AMDNLoc framework across various scenarios generated through ray tracing, utilizing different datasets. The selected scenarios include DeepMIMO O1 blockage, WAIR-D 00247, and WAIR-D 00743. The results, compiled after 400 epochs, are presented in Tab. \ref{tab:multidataset}. A key observation is that the segment-specific linear array demonstrates greater improvements in accuracy, especially in scenarios with an increased number of buildings compared to those without the array.

Furthermore, in datasets with a higher density of MTs, such as the DeepMIMO O1 blockage scenario where the spatial distribution of sample MTs is more uniform and compact, the use of ADCAM as fingerprints yields results remarkably close to those obtained using AOA, AOD, distance, and gain. Notably, despite the augmented complexity of these datasets, the optimal MSE consistently remains within a 2-meter range. This consistent performance, even under varying conditions and increased dataset complexity, highlights the robustness and adaptability of the proposed AMDNLoc framework across diverse scenarios and MT spatial distributions, and underscores its substantial potential for future applications in next-generation 
multiple access environments.

\begin{table}[h!]
\centering
 \caption{MSE of multiple scenarios and datasets with UPA array.}
 \label{tab:multidataset}
 \begin{tabular}{c c c c} 
 \toprule
 Methods & \makecell[c]{DeepMIMO \\ O1 blockage} & \makecell[c]{WAIR-D \\ 00247} & \makecell[c]{WAIR-D \\ 00743}\\ [0.2ex] 
 \midrule
 Res\_CFRADP & 1.66 & 2.23 & 4.21 \\ 
 Res\_multi\_CFRADP & 1.11 & 1.54 & 1.99  \\
 Res\_multi\_CFRperfectADP & 0.92 & 1.11 & 1.25  \\
 \bottomrule
 \end{tabular}
\end{table}

\subsubsection{Multiple Feature Extractors}
\begin{table}[h!]
\centering
 \caption{MSE and running time of different feature extractors.}
 \begin{tabular}{c c c c c} 
 \toprule
 Epochs & Items & DenseNet121 & ResNet18 & EfficientformerV2 \\
 \midrule
 \multirow{2}{*}{150} & MSE(m) & 2.44 & 2.54 & 2.58 \\  & Time(s) & 5988 & 2192 & 8050 \\
 \hline
 \multirow{2}{*}{250} & MSE(m) & 2.27 & 2.24 & 2.28 \\  & Time(s) & 9946 & 3655 & 13415 \\
 \bottomrule
 \end{tabular}
 \label{tab:resultsextractor}
 \vspace{-3mm}
\end{table}
We conduct an evaluation to assess the adaptability of the AMDNLoc framework with various neural networks, using different feature extractors such as DenseNet121 and EfficientformerV2, as outlined in Tab. \ref{tab:resultsextractor}. Through this comparative analysis, we found that ResNet18 offers the most effective balance between time efficiency and accuracy. DenseNet18, while achieving better accuracy, is slower than ResNet18 due to its use of dense connections instead of skip connections.

While lightweight Vision Transformer (ViT) models are generally known for outperforming ResNet models in certain applications \cite{chen2021vision}, in our specific context, ResNet18 demonstrates superior speed and performance compared to the latest lightweight ViT models. This can be attributed to the fact that DCNN has historically been more adept at extracting features from CSI data \cite{wang2016csi_fingerprint}. Our pre-classification function effectively organizes the data, ensuring similar fingerprint distributions within each class, which is conducive to feature extraction by ResNet. 
However, the primary challenge with ViT lies in handling inductive bias. In theory, our pre-classification should provide a favorable inductive bias for ViT, potentially leading to better results. Nonetheless, ViT’s approach of decomposing images into batches for network training poses a limitation. Given the high similarity within each batch of CFR data, ViT struggles to effectively manage this translation and similarity, leading to suboptimal outcomes. Despite this, we believe that further exploration and modifications of the existing ViT model could yield improvements, presenting a promising avenue for future research.


\subsubsection{From Only NLOS to NLOS and LOS}

\begin{figure}\vspace{-0mm}
	\begin{center}
		\centerline{\includegraphics[width=0.45\textwidth]{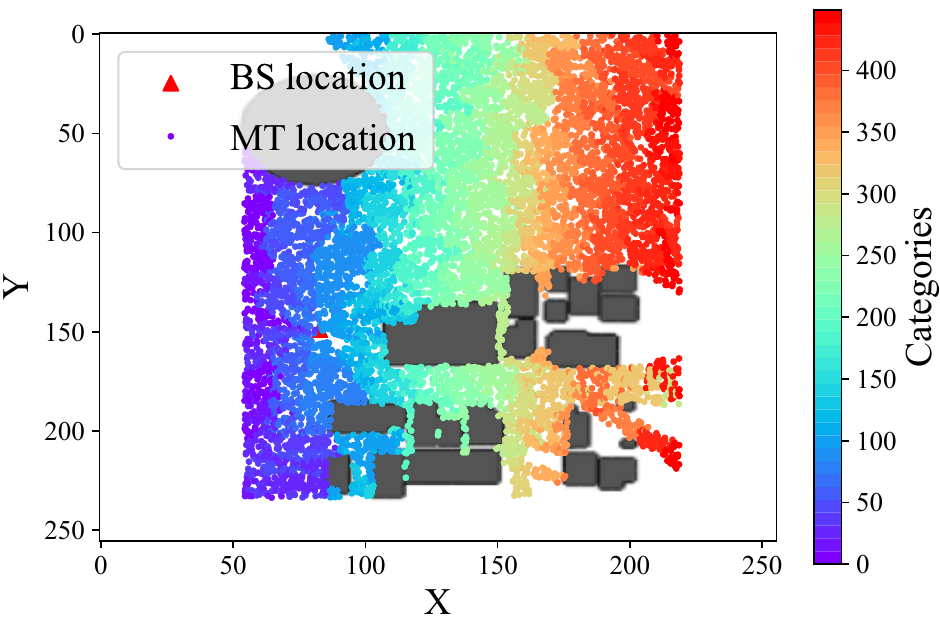}}  \vspace{-0mm}
		\caption{NLOS and LOS pre-classification in combining domain.}
		\label{fig:bothvisual} \vspace{-9mm}
	\end{center}
\end{figure}

We have expanded our study to include scenarios that encompass both NLOS and LOS conditions. The classification areas under these combined conditions are visualized in Fig. \ref{fig:bothvisual}. Similar to the observations in Fig. \ref{fig:visulizaion1.3}, our AMDNLoc framework demonstrates effective intra-group continuity and inter-group discrimination. Notably, it achieves an accuracy of less than 2 meters.

It is important to highlight that CSI is different in NLOS and LOS scenarios. Consequently, existing research mainly focuses on either mitigating or identifying NLOS conditions prior to localization \cite{darbinyan2023ml, li2017nlos}. However, we present a novel perspective by integrating both NLOS and LOS conditions to a certain extent. The effectiveness of AMDNLoc in capturing the distinct feature distributions of channel information suggests its broader applicability that is not limited to multi-points fingerprint localization scenarios.

\section{Conclusion}\label{sec:conclu}
In the paper, we proposed a novel multi-sources information fusion learning framework that makes full use of the multi-path features across the frequency, power, angle and delay domain as fingerprints to tackle the inherent heterogeneity issue of fingerprint distributions. 
We developed a two-stage matched filter for PFCFR corresponding to the distribution of CFR, and fused it with ADCAM classification region after centroid-based clustering method. 
Furthermore, a segment-specific linear classifier mechanism sharing the same feature extractor was utilized to build regression relationship between fingerprint and locations after eliminating the negative effect of the region covariant. 
Our numerical experiments demonstrated that AMDNLoc achieves SOTA results, outperforming traditional convolutional neural networks in datasets such as WAIR-D and DeepMIMO. These findings highlighted AMDNLoc’s exceptional accuracy, interpretability, adaptability, and scalability, marking a significant advancement in the field of wireless communication and localization.

\small
\bibliographystyle{IEEEtran}
\bibliography{bib}
\vspace{12pt}

\end{document}